\documentclass[aps,prb,superscriptaddress,preprintnumbers,
showpacs,twoside,twocolumn,amsmath,amssymb]{revtex4}
\usepackage{txfonts}
\usepackage{amsfonts}

\usepackage{graphicx}     
\usepackage{bm}

\allowdisplaybreaks

\begin{document}
\title{Neutron and ARPES Constraints on the Couplings of the \\
       Multiorbital Hubbard Model for the Pnictides}
\author{Qinlong Luo}
\affiliation{Department of Physics and Astronomy, University of
Tennessee, Knoxville, Tennessee 37996, USA} \affiliation{Materials Science
and Technology Division, Oak Ridge National Laboratory, Oak Ridge,
Tennessee 32831, USA}
\author{George Martins}
\affiliation{Physics Department, Oakland University, Rochester, Michigan 48309, USA}
\author{Dao-Xin Yao}
\affiliation{Department of Physics and Astronomy, University of
Tennessee, Knoxville, Tennessee 37996, USA} \affiliation{Materials Science
and Technology Division, Oak Ridge National Laboratory, Oak Ridge,
Tennessee 32831, USA}\affiliation{State Key Laboratory of Optoelectronic Materials and
Technologies, School of Physics and Engineering, Sun Yat-Sen University, Guangzhou 510275, China}
\author{Maria Daghofer}
\affiliation{IFW Dresden, P.O. Box 27 01 16, D-01171 Dresden, Germany}
\author{Rong Yu}
\affiliation{Department of Physics and Astronomy, Rice University, Houston, Texas 77005, USA}
\author{Adriana Moreo}
\affiliation{Department of Physics and Astronomy, University of
Tennessee, Knoxville, Tennessee 37996, USA} \affiliation{Materials Science
and Technology Division, Oak Ridge National Laboratory, Oak Ridge,
Tennessee 32831, USA}
\author{Elbio Dagotto}
\affiliation{Department of Physics and Astronomy, University of
Tennessee, Knoxville, Tennessee 37996, USA} \affiliation{Materials Science
and Technology Division, Oak Ridge National Laboratory, Oak Ridge,
Tennessee 32831, USA}
\date{\today}

\begin{abstract}
The results of neutron scattering and angle-resolved photoemission experiments for the
Fe-pnictide parent compounds, and their metallic nature, are shown to 
impose severe constraints on the range of values that can be considered ``realistic''
for the intraorbital Hubbard repulsion $U$ and Hund coupling $J$ in multiorbital Hubbard models treated in the mean-field
approximation. Phase diagrams for three- and five-orbital models are here provided, and
the physically realistic regime of couplings is highlighted, to guide future theoretical
work into the proper region of parameters of Hubbard models. In addition, using the random phase
approximation, the pairing tendencies in these realistic coupling regions are
investigated. It is shown that the dominant spin-singlet pairing channels 
in these coupling
regimes correspond to nodal superconductivity, with strong competition between 
several states that belong to different irreducible representations. This is compatible with experimental bulk measurements 
that have reported the existence of nodes in several Fe-pnictide compounds.
\end{abstract}


\maketitle

\section{Introduction}

An exciting new area of research has recently opened with the 
discovery of superconductivity in the Fe pnictides.\cite{Fe-SC,chen1,chen2,chen3,55,ren2,david-j}
These materials present many similarities with the high temperature
superconductors based on copper, but also several differences.
Both families have layered structures suggesting that theoretical studies in 
two dimensional lattices should be able to capture the essence of the pairing mechanism. 
In the undoped limit, both types of compounds are magnetic, with wavevector $(\pi,0)$ in
the case of the pnictides\cite{dai-lynn,andy} and $(\pi,\pi)$ for the cuprates, in the notation of the square
lattice defined by Fe or Cu. However, a crucial difference is that the undoped parent 
compound is an insulator for the cuprates, while it is a (bad) metal for the pnictides. 
This fact already suggests that the regime of a large Hubbard coupling $U$, widely used in
the context of the cuprates, may not be appropriate for a theoretical description of the pnictides.

Adding to the complexity of this problem, it is also clear that
a theoretical study of pnictides cannot rely on just one orbital, as in the case of the cuprates,
but it needs a multiorbital approach.\cite{Cao08} In fact, to study the magnetic and superconducting properties of the pnictides, a considerable effort has already started using multiorbital model Hamiltonians.\cite{Raghu08,Kuroki,Daghofer08,Eremin,Si,Hu,Lorenzana,Yuetal08,Schmalian,Moreo09,Graser08,Graser10,chen102,calderon,calderon2,wku,plee,laad,0910.2707,moreo09bis,ThreeOrbital,constraints,newpaper} 
Having to consider multiple orbitals severely restricts the available tools to carry
out unbiased computer-based investigations of Hubbard multiorbital models. As a consequence, several studies
have been restricted to mean-field approximations. Fortunately, this is not a drastic limitation for the case of the
undoped systems, since similar mean-field approximations for the cuprates 
are known to capture qualitatively the essence
of the magnetic states.\cite{schrieffer}
In fact, recent mean-field-based efforts for the pnictides have already reported
the presence of a state that is simultaneously metallic and
magnetic.\cite{Yuetal08} However, the study of the intermediate coupling regime, where
the magnetic-metallic state was found,
establishes a considerable challenge to theory
since an intermediate region of couplings is often more difficult to analyze than either
extreme of large or small $U$. In addition, a multiorbital Hubbard 
approach needs at least two couplings: the on-site intra-orbital Hubbard repulsion $U$ and the
on-site Hund coupling $J$. (A third parameter, the on-site inter-orbital repulsion $U'$, is then defined by the well-known relation
$U = U' +2J$ arising from symmetries in orbital space.~\cite{oles83}) Having two couplings 
increases further the complexity of the
analysis and the comparison between different approaches since there is at present no universally accepted
range of $U$ and $J$ that is considered realistic by the community of experts. In fact, it 
would be quite desirable to restrict the values of the $U$ and $J$ couplings used in the literature
to a much narrower range, where qualitative agreement with experiments is observed.

In this publication, our goal is to use experimental neutron scattering and photoemission data for the
undoped pnictide parent compounds, supplemented by their 
well-known metallic properties, to establish lower and
upper bounds on the couplings $U$ and $J$ of multiorbital Hubbard models. 
By focusing on a more restricted set of
couplings, our results will guide future theoretical efforts into a realistic regime for the pnictides.
Our calculations are based on the previously used and tested mean-field 
approximation,\cite{Yuetal08} allowing us to
calculate a variety of observables that are then compared against experimental results to establish the
proper ranges for $U$ and $J$. 

To carry out this theory-experiment comparison the focus here is on the results of two powerful
experimental techniques. One of them is
neutron scattering. In these
experiments, a magnetic order with wavevector $(\pi,0)$ (in the Fe square-lattice notation)
was observed at low temperatures.\cite{dai-lynn,andy} 
This particular wavevector can be theoretically
accommodated by merely using proper tight-binding model Hamiltonians that reproduce the
well-known paramagnetic Fermi surface (with hole and electron pockets). More
important for our purposes is the actual value of the ordered moment
for the iron spin, $\mu_{\rm Fe}$.
Table 2 of Ref.~\onlinecite{dai-lynn} and Table 2 of Ref.~\onlinecite{andy} provide a summary of the
neutron scattering experimental values for $\mu_{\rm Fe}$: within the 1111 and 122 families
they range from 0.25 for NdOFeAs to $\sim 1$ for SrFe$_2$As$_2$ and BaFe$_2$As$_2$ 
(the 11 family has substantially larger
values of $\mu_{\rm Fe}$, but their ordering wavevectors are different, 
indicating that the 11 compounds
require a special discussion beyond the scope of this work). Since the mean-field investigations
reported here focus on the mean-field solution corresponding to wavevector $(\pi,0)$, the
theoretical value of $\mu_{\rm Fe}$ can be obtained when 
varying the couplings in the Hubbard model,
and a range compatible with experiments can be found. Not surprisingly, considering the
(bad) metallic character of these compounds, the range of interest will be shown to correspond
to intermediate values of the on-site repulsion $U$, as also discussed before.\cite{Yuetal08,calderon2} However,
our investigations below show that sizable constraints over the Hund coupling $J$, not explored before, 
can also be obtained.

Another powerful experimental technique that will help us to establish constraints on
the couplings of the Hubbard models is Angle Resolved Photoemission (ARPES). Applying this technique,
several reports describing the Fermi surface of the pnictides have been presented, including their temperature
evolution. Here, the vast ARPES literature will not be comprehensively reviewed but the focus will
be on the results for the undoped compounds in the spin density wave (SDW) regime. In particular, it has been reported
by several groups that in this magnetically ordered 
state the Fermi surface presents extra ``features''
in the vicinity of the $\Gamma$-point hole pockets that are 
not present in band-structure calculations for the paramagnetic state. 
These extra features, mainly found for the 122 compounds, 
are described as the existence of ``satellite''
pockets caused by $V$-shaped bands 
(or by $V$-shaped band crossings).~\cite{photo2,photo3,photo5,photo6,photo1,photo7,photo4} 
Our focus will be on trying
to reproduce qualitatively with mean-field approximations these type of satellite
extra features that ARPES investigations have systematically reported in
the magnetic state. Note that near the original electron pockets at $(\pi,0)$ and $(0,\pi)$ there are also
several reports of interesting modifications  of  the  
paramagnetic band structure when at low temperatures. However, in our opinion, the ARPES 
analysis of the modifications to the 
band structure electron pockets appears less robust due to the presence of considerable 
noise in the raw data, more than for the case of the hole pockets. For this reason, the focus here will be
on the existence of satellite 
pockets at low temperatures near the original
$\Gamma$-point hole-pockets. Note also that these satellite features 
have been given an electron-like character in some publications\cite{photo5}
 but considering the uncertainties in ARPES studies, mainly caused by
the considerable backgrounds, our analysis below will focus on finding
regions in the $U$-$J/U$ plane where {\it any} kind of extra pocket is
induced by the magnetic order close to the original 
$\Gamma$-point hole-pockets, regardless of whether its character is electron-
or hole-like.

While finding constraints on  $U$ and $J$ in multiorbital Hubbard models is already
interesting, 
our effort here continues with the analysis of the pairing states that are obtained
in those ``realistic'' $U$-$J/U$ regions, via the use of the Random Phase Approximation (RPA).\cite{Graser08,Graser10} Exploring
pairing tendencies
is important since there is a growing controversy in the Fe pnictides investigations with
regards to the symmetry of the superconducting order parameter. While a variety of photoemission
experiments report the absence of nodes,\cite{y1,y2,y3,y4,y5,y6} 
at least clearly with regards to the $\Gamma$-point hole pockets, other bulk experiments report results 
compatible with a nodal superconducting state.\cite{n1,n2,n3,n4,n5,n6,n7,n8,n9,n10,n11,n12} 
This controversy
has not been resolved and it is one of the most important open problems in pnictides. Our
investigations show that in the realistic $U$-$J/U$ regimen identified here, for both three-
and five-orbital models, the RPA dominant superconducting state is {\it nodal} (this includes the
$A_{ 1g}$ state with zeros (or near zeros) of the order parameter). Moreover, 
within RPA there is a clear competition 
between a variety of states belonging
to different irreducible representations suggesting that
different pnictides may have different symmetries with regards to  the 
superconducting order parameter. All these competing states are
nodal or quasi-nodal, at least within the limitations of our approximations.

The organization of this manuscript is the following. First, the models and technique used
are presented in Sec.~\ref{model}. This is followed  in Sec.~\ref{three} 
by the search for a realistic $U$-$J/U$
regime for the three-orbital model. The case of five orbitals is presented
in Sec.~\ref{five}. The dominant RPA pairing tendencies in the realistic regime are presented
in Sec.~\ref{rpa}. Finally, conclusions are provided in Sec.~\ref{conclusions}.

\section{Models and Techniques}
\label{model}

\subsection{The model Hamiltonians}

In this effort, the three-orbital Hubbard model introduced in Ref.~\onlinecite{ThreeOrbital} will
be used first. This model is purely based on the $d$ electrons of Fe and it 
considers only the
three orbitals $d_{ xz}$, $d_{ yz}$, and $d_{ xy}$, widely believed to be the most relevant
orbitals at the Fermi surface for the pnictides.
The reader is referred to the original publication\cite{ThreeOrbital} for a full description
of the model and its band structure (which is in good agreement with ab-initio calculations).
In momentum space, the model includes a tight-binding term defined as:
\begin{eqnarray}\label{E.H0k}
H_{\rm TB}(\mathbf{ k}) &=& \sum_{\mathbf{ k},\sigma,\mu,\nu} T^{\mu,\nu}
(\mathbf{ k})
d^\dagger_{\mathbf{ k},\mu,\sigma} d^{\phantom{\dagger}}_{\mathbf{ k},\nu,\sigma}~,
\end{eqnarray}
with
\begin{eqnarray}
T^{11} &=& 2t_2\cos  k_x +2t_1\cos  k_y +4t_3 \cos  k_x
\cos  k_y-\mu, \label{eq:t11}\\
T^{22} &=& 2t_1\cos  k_x +2t_2\cos  k_y +4t_3 \cos  k_x
\cos  k_y-\mu, \label{eq:t22}\\
T^{33} &=& 2t_5(\cos  k_x+\cos  k_y) \nonumber\\
       & & +4t_6\cos  k_x\cos  k_y -\mu +\Delta_{xy},
\label{eq:t33} \\
T^{12} &=& T^{21} =4t_4\sin  k_x \sin  k_y, \label{eq:t12}\\
T^{13} &=& \bar{T}^{31} = 2it_7\sin  k_x + 4it_8\sin  k_x \cos  k_y,
\label{eq:t13} \\
T^{23} &=& \bar{T}^{32} = 2it_7\sin  k_y + 4it_8\sin  k_y \cos  k_x\;,
\label{eq:t23}
\end{eqnarray}
where a bar on top of a matrix element denotes the complex conjugate. $\mu$ and 
$\nu$ range from 1 to 3 and label the orbitals $d_{ xz}$ (1), $d_{ yz}$
(2), and $d_{ xy}$ (3).
Since the
Hamiltonian for a one-iron unit cell is here considered, then ${\bf k}$ runs
within the corresponding extended Brillouin zone (BZ) $-\pi<k_x,k_y\le\pi$. The actual values
for the hopping amplitudes are in Table I.

\begin{table}
\caption{Parameters for the tight-binding portion of the three-orbital model
  used here. The overall energy unit is
  electron volts.\label{tab:hopp3}}
 \begin{tabular}{|ccccccccc|}\hline
$t_1$ & $t_2$ & $t_3$ & $t_4$ & $t_5$ & $t_6$ & $t_7$ & $t_8$ &
   $\Delta_{xy}$\\
\hline
  0.02   &0.06    &0.03   &$-0.01$&$0.2$ & 0.3 & $-0.2$ & $-t_7/2$&
  0.4\\ \hline
 \end{tabular}
\end{table}

The Coulombic interacting portion of the three-orbital Hamiltonian is given by:
\begin{equation}\begin{split}  \label{eq:Hcoul}
  H_{\rm int}& =
  U\sum_{{\bf i},\alpha}n_{{\bf i},\alpha,\uparrow}n_{{\bf i},
    \alpha,\downarrow}
  +(U'-J/2)\sum_{{\bf i},
    \alpha < \beta}n_{{\bf i},\alpha}n_{{\bf i},\beta}\\
  &\quad -2J\sum_{{\bf i},\alpha < \beta}{\bf S}_{\bf{i},\alpha}\cdot{\bf S}_{\bf{i},\beta}\\
  &\quad +J\sum_{{\bf i},\alpha < \beta}(d^{\dagger}_{{\bf i},\alpha,\uparrow}
  d^{\dagger}_{{\bf i},\alpha,\downarrow}d^{\phantom{\dagger}}_{{\bf i},\beta,\downarrow}
  d^{\phantom{\dagger}}_{{\bf i},\beta,\uparrow}+h.c.),
\end{split}\end{equation}
where $\alpha,\beta=1, 2, 3$ denote the orbitals, ${\bf S}_{{\bf i},\alpha}$
($n_{{\bf i},\alpha}$) is the spin (electronic density) of orbital $\alpha$ at site
${\bf i}$ (this index labels sites of the square lattice defined by the irons), 
and the relation $U'=U-2J$ between the Kanamori parameters
has been used.\cite{manga} 
The first two terms give
the energy cost of having two electrons located in  the same orbital or in
different orbitals, both at the same site, respectively. The second line contains the Hund's
rule coupling that favors the ferromagnetic (FM) alignment of the spins in
different orbitals at the same lattice site. 
The ``pair-hopping'' term is in the third line and its
coupling is equal to $J$ by symmetry.
Note that the values used for $U$ and $J$ can be substantially smaller
than the atomic ones, because the interactions may be screened by
bands not included in the Hamiltonian. The Coulombic interaction terms 
introduced above have
been used and discussed in several previous 
publications~\cite{Daghofer08,Yuetal08,Moreo09,ThreeOrbital,calderon2}
where more details can be found by the readers. All energies are provided here in
electron volts. As shown in Ref.~\onlinecite{ThreeOrbital}, the electronic density of
relevance for this model is $n=4$ to reproduce the expected Fermi surface 
in the paramagnetic regime.

In the present investigation, two five-orbital models (also based only on the $d$ electrons of Fe) 
have also been used, at electronic density $n=6$. By 
supplementing the three-orbital model by more complicated five orbital versions, our main
goal is to
verify the self-consistency of our approach. In other words, if the many models, with
similar Fermi surfaces by construction, would give quite different ranges of couplings for
the compatibility with neutron and ARPES results, then this would raise concerns about
the entire calculation. It turns out that, as shown below, the $J/U$ and $U$ ranges that
are found to be physically reasonable are similar in all cases, 
demonstrating that our approach
is self-consistent. With regards to the specific five-orbital
models used here, the tight-binding parameters of one of them 
are in the Appendix, while another set of hoppings is from Ref.~\onlinecite{Graser08}.
At $U=0$, all these
models provide 
a Fermi surface (see below) that compares well with experiments and band structure calculations for the 
122 compounds. The Coulombic interactions for five-orbitals are the obvious generalization of the terms used
for three-orbitals.

\subsection{The mean-field approximation}

To study the ground state properties of the models introduced before, 
a mean-field approximation will be applied. This approximation was already
presented in previous publications,\cite{Yuetal08,calderon2} but it is here also
discussed for completeness. The simple standard assumption of considering only the
mean-field values for the diagonal operators is followed:\cite{nomura}

\begin{eqnarray}\label{E.MFA}
\langle d^\dagger_{{\bf i},\mu,\sigma} d_{{\bf j},\nu,\sigma'}\rangle =
\left(n_\mu+\frac{\sigma}{2}\cos(\mathbf{q}\cdot\mathbf{r}_{\bf i})m_\mu\right)
\delta_{\bf ij}\delta_{\mu\nu}\delta_{\sigma\sigma'},
\end{eqnarray}
where $\mathbf{q}$ is the ordering wavevector of the magnetic
order. $n_\mu$ and $m_\mu$ are mean-field parameters (to be determined self-consistently) 
describing the
charge density and magnetization of the orbital $\mu$, respectively. 
The rest of the
notation is standard. Applying
Eq.~(\ref{E.MFA}) to $H_{\rm int}$, the mean-field Hamiltonian in
momentum space can be written as

\begin{eqnarray}\label{E.HMF}
H_{\rm MF} = H_{\rm TB} + C + \sum_{\mathbf{k},\mu,\sigma}
\epsilon_\mu d^\dagger_{\mathbf{k},\mu,\sigma}
d_{\mathbf{k},\mu,\sigma}\nonumber\\
+ \sum_{\mathbf{k},\mu,\sigma} \eta_{\mu,\sigma}
 (d^\dagger_{\mathbf{k},\mu,\sigma} d_{\mathbf{k+q},\mu,\sigma} +
d^\dagger_{\mathbf{k+q},\mu,\sigma} d_{\mathbf{k},\mu,\sigma}),
\end{eqnarray}
where $\mathbf{k}$ runs over the extended first BZ, $H_{\rm TB}$ is
the hopping term in Eq.~(\ref{E.H0k}), the constant $C$ is
\begin{eqnarray}
C=&-&NU\sum_{\mu}\left(n^2_\mu-\frac{1}{4}m^2_\mu\right)
- N(2U'-J)\sum_{\mu\neq\nu}n_\mu n_\nu \nonumber \\
&+& \frac{NJ}{2} \sum_{\mu\neq\nu} m_\mu m_\nu, \nonumber
\end{eqnarray}
$N$ is the number of sites, and the following definitions were introduced
\begin{eqnarray}
\epsilon_\mu = Un_\mu + (2U'-J)\sum_{\nu\neq\mu}
n_\nu, \\
\eta_{\mu,\sigma} =
-\frac{\sigma}{2}\left(Um_\mu+J\sum_{\nu\neq\mu}m_\nu\right).
\end{eqnarray}

The mean-field Hamiltonian can be numerically solved for a fixed
set of mean-field parameters using standard
library subroutines. $n_\mu$ and $m_\mu$ are obtained 
self-consistently by minimizing the energy via an iterative process. 
During the iterations
$\sum_\mu n_\mu$=$n$ was enforced at each step, such that
the total charge density is a constant (4 for the three-orbital model,
and 6 for the five-orbital models). The reader should assume that these
are the electronic densities used for these models throughout the manuscript,
both in the mean-field approximation and for the RPA approximation as well. 
Note also that 
the numerical solution of the mean-field Hamiltonian immediately allows for the
calculation of the band structure, density of states (DOS), and
magnetization ($m=\sum_\mu m_\mu$) at the ordering wavevector
$\mathbf{q}$. Moreover, the photoemission
spectral function can also be calculated, as explained in
Ref.~\onlinecite{Yuetal08}.

\section{Results for the Three-Orbital Model}
\label{three}

Our discussion of results starts with the three-orbital model.
Some aspects of this discussion have been briefly mentioned in other publications,
thus references to those previous efforts are provided where appropriate.

\subsection{Comparison with neutron scattering experiments}

\begin{figure}[ht]
\begin{center}
\includegraphics[clip,width=80mm,angle=0]{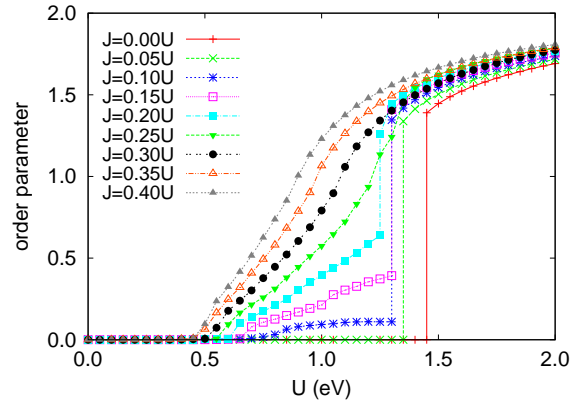}
\caption{Mean-field order parameter at wavevector $(\pi,0)$ 
vs. $U$ (in eV units) for the three-orbital model discussed
in the text, and parametric with the values of $J/U$ indicated.} \label{mU3}
\end{center}
\end{figure}

Figure~\ref{mU3} contains the mean-field order parameter ($m$) at wavevector $(\pi,0)$ vs.
$U$. The plotted
values for $m$ arise from the numerical solution of the mean-field
equations discussed in the previous section.
For small values of $J/U$, such as
$0.00$ and $0.05$, in Fig.~\ref{mU3} $m$ discontinuously jumps from zero to a robust value at
a critical $U$. 
While such a discontinuity is observed in all models
discussed in this paper, its origin is not universal. 
It is
caused by a metal-insulator transition in a four-band model, see
Sec.~II.C.4 of Ref.~\onlinecite{Yuetal08}. 
In the three-orbital model,
the discontinuity only coincides with the opening of a gap for smaller 
$J/U\lesssim 0.15 $, and is
rather marked by the sudden onset of strong orbital
order, with a close competition between substantial alternating and nearly perfect ferro-orbital order.\cite{ThreeOrbital} For larger  a$0.15 \lesssim J/U \lesssim 0.22$, states with both types of orbital order 
can remain metallic for a small range of $U$ just above the onset 
of strong orbital order, but the FS is qualitatively very different from ARPES
results, e.g. it does not feature any sign of hole pockets around the
$\Gamma$ point.\cite{ThreeOrbital}
At or rapidly after the critical $U$, the density of states develops a gap (not shown), signaling insulating behavior, in contradiction with the 
experimentally observed (bad) metallic character of the undoped pnictides. 
Moreover, as a consequence of the discontinuity, the order parameter $m$ in the range of small $J/U$ never
reaches the realistic values for pnictides reported in neutron scattering experiments, i.e. [0.25,1.0] for $(\pi,0)$ magnetic order.
Thus, these results for small $J/U$
start illustrating one of the main messages of this publication,
namely that within the mean-field approximation used in our effort the request of
qualitative agreement with the experimental properties of the undoped pnictides imposes
severe constraints on the values of $U$ and $J/U$ for the multi-orbital Hubbard models.
In particular, it is clear that $J/U = 0.00$ and $0.05$ do not seem 
physically appropriate to describe the pnictides.

As $J/U$ increases further, $m$ now develops (becomes nonzero) at an earlier
critical value of $U$, allowing for a proper description of materials with weak
magnetic order parameters such as the ``1111'' family. Note that for
$J/U = 0.10$, $0.15$, and $0.20$ , a
discontinuity is still present in $m$ vs. $U$, so not all values of the order parameter $m$
are possible, while for larger $J/U$s the $m$ curves are no longer discontinuous.
$J/U = 0.50$ is the largest ratio that should be considered to avoid
a negative $U'$ due to the relation $U=U'+2J$. 
Thus, adding this information to the results for
$m$ and its comparison with neutron scattering, the
proper range of $J/U$ couplings naively becomes $\sim [0.10, 0.50]$, with $U$ larger than
the first critical value where $m$ develops. However, if in addition it is considered that $J$
should be smaller than $U'$, then this reduces the range further 
to $[0.10,0.33]$, since $J = U'$ at $J/U=1/3$.

\subsection{Comparison with ARPES experiments}

As discussed in the Introduction, another experimental source of information
that can be used to reduce the allowed range of couplings in the Hubbard model is provided
by the ARPES results for undoped pnictides. As previously mentioned, a common
generic feature of several ARPES experiments at low temperatures in the SDW
phase is the development of ``extra'' features (pockets) near the original
$\Gamma$-point 
hole pockets of the noninteracting limit. To search for these features, within our
mean-field approximation the one-particle spectral function $A({\bf k},\omega)$ has
been calculated, and the Fermi surface results have been analyzed in a wide range
of $U$ and $J/U$, using a $\delta$-function broadening 0.025 eV.

\begin{figure}[ht]
\begin{center}
\includegraphics[clip,width=50mm,angle=0]{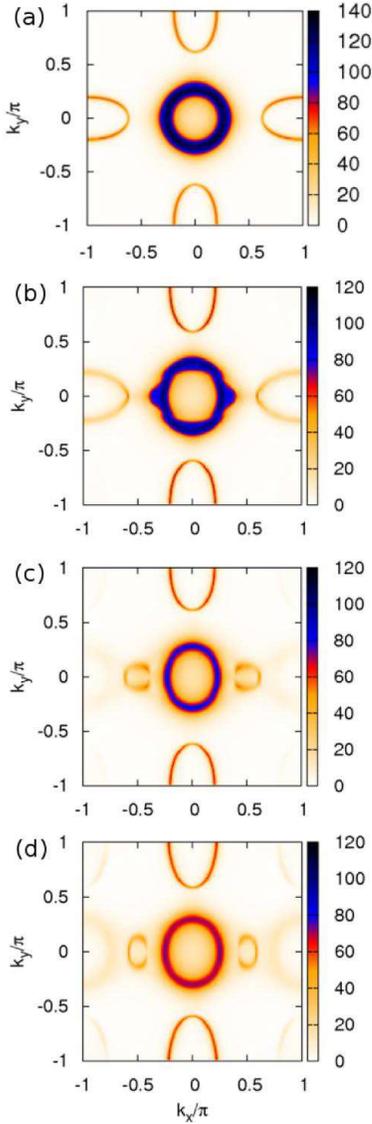}
\caption{ Unfolded mean-field Fermi surfaces for the three-orbital model 
corresponding to ($m$=order parameter)
(a) $U=0$, as reference (there are two hole pockets at $\Gamma$ but they appear
merged due to the broadening used for plotting);  
(b) $J/U=0.33$, $U=0.6$, $m=0.2$;
(c) $J/U=0.20$, $U=1.0$, $m=0.4$; and 
(d) $J/U=0.25$, $U=1.05$, $m=0.6$.
} \label{FS3}
\end{center}
\end{figure}

Shown in Fig.~\ref{FS3} are representative results of our ARPES calculations 
(see also Refs.~\onlinecite{ThreeOrbital,newpaper}). The focus is on
a range of $J/U$ and $U$ where mean-field ARPES contains a $\Gamma$-point hole pocket, as in
the original $U=0$ bandstructure (presumably corresponding to the high
temperature non-magnetic regime as well), 
and in addition ``satellite'' pockets as in ARPES, in
between the original hole and electron pockets along the $k_x$ axis. 
The results have not been folded, but
are representative of a single-domain spin order wavevector, in this case $(\pi,0)$, and
with only one Fe per unit cell. Figure~\ref{FS3} shows representative cases where
these satellites are clearly present [the satellite pockets 
tend to be electron-like for small $U$, switching to
hole-like for slightly larger, but still realistic, values of $U$ (for more details 
see Ref.~\onlinecite{ThreeOrbital})]. 
The corresponding values of $J/U$ and $U$ are
indicated in the figure caption. These physically acceptable Fermi surfaces from the ARPES
perspective are found in the same approximate range of couplings as those selected
from the $m$/neutrons perspective, as discussed above. It has been suggested
before that a reduced ordered magnetic
moment goes together with realistic $A({\bf k},\omega)$ in numerical
approaches like density-functional theory~\cite{photo5} and the present mean-field
scheme,~\cite{newpaper} and we see here that this is not accidental, but that the analysis 
of both neutrons and ARPES are mutually consistent
over a larger parameter range, as well as for a variety of models, see
Sec.~\ref{five}.
The four panels shown are qualitatively similar and 
further refinements in the so-called ``physical region'' (see Fig.~\ref{PhaseDiagram3}) 
will need better tools for calculations and more accurate ARPES experiments.

\begin{figure}[ht]
\begin{center}
\includegraphics[clip,width=70mm,angle=0]{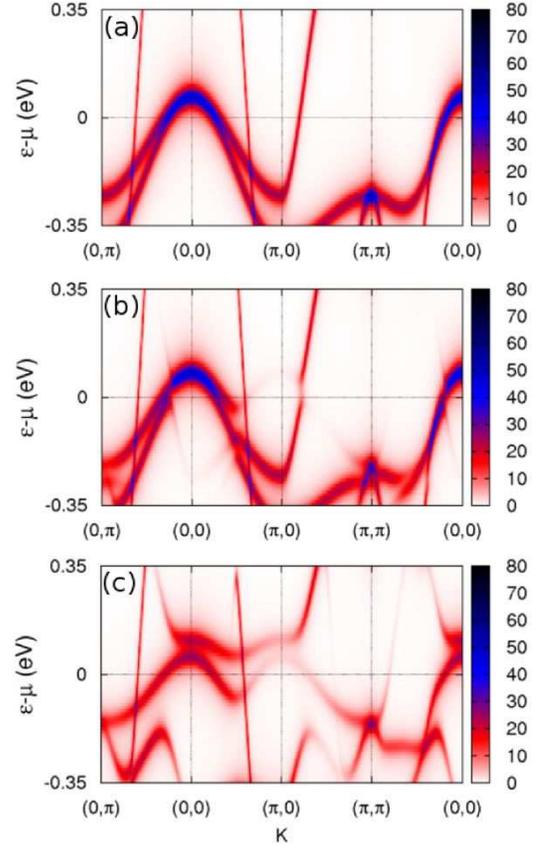}
\caption{Unfolded band structure mean-field results for the three-orbital model and cases 
(a) $U=0$, as reference;  
(b) $J/U=0.33$, $U=0.6$, $m=0.2$; (c) $J/U=0.25$, $U=1.05$, $m=0.6$.
Panels (b) and (c) show a $V$-shaped pocket in between the 
$(0,0)$ and $(\pi,0)$ points. The scale used (arbitrary units) is on the 
right of the panels.} \label{bands3}
\end{center}
\end{figure}

In Fig.~\ref{bands3}, some of the full spectral functions are shown, and compared with the $U$=0
case. The appearance of $V$-shaped features, that induce 
the presence of satellite pockets,
is clear in these figures. These mean-field
results for $A({\bf k},\omega)$ are qualitatively consistent with ARPES
experiments for the pnictides 
that have reported similar $V$-shaped branches.\cite{photo6}

The results of Fig.~\ref{FS3} are in qualitative 
agreement with experiments, as already remarked 
in Ref.~\onlinecite{newpaper}. 
Moreover, our comprehensive analysis of $A({\bf k},\omega)$ has
shown that these features
do not appear in other regions of the $U$-$J/U$ phase diagram. For instance, before the
critical $U$ where $m$ develops from zero 
there are no satellite pockets, since they arise from
the nonzero magnetic order and nesting effects. In the other extreme of $U$ couplings
larger than those used in Fig.~\ref{FS3}, 
the Hubbard model simply becomes insulating (as
discussed before in Ref.~\onlinecite{Yuetal08}), and there is no
longer a Fermi surface.

\subsection{Summary phase diagram for three orbitals}

\begin{figure}[ht]
\begin{center}
\includegraphics[clip,width=90mm,angle=0]{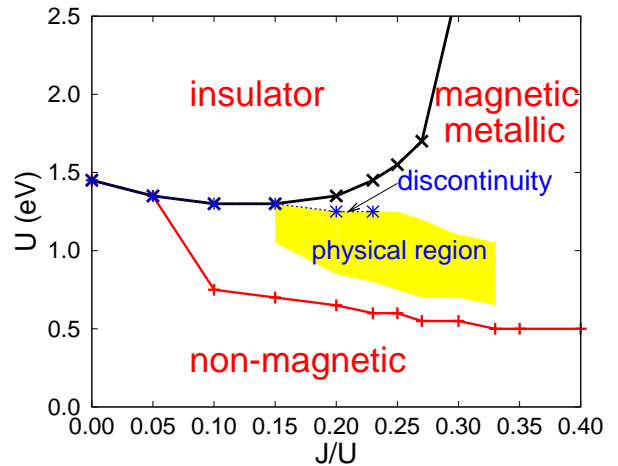}
\caption{Phase diagram for the three-orbital model obtained with the mean-field
approximation described in the text. The ``physical region'' in yellow is the
regime of couplings found to be compatible with neutron and photoemission experiments.
The ``non-magnetic'' region corresponds
to a regime where the state has a zero order parameter. In the ``insulator'' region,
there is no Fermi surface and the state is insulating. The ``discontinuity'' label
corresponds to the discontinuous jump in the order parameter shown in Fig.~\ref{mU3}. The
entire ``magnetic metallic'' regime could in principle have been compatible with
experiments, but only in the yellow highlighted region is that $m$ is sufficiently
small-intermediate in value and the Fermi surface has satellite pockets
near the $\Gamma$-point hole pockets. 
%
} \label{PhaseDiagram3}
\end{center}
\end{figure}

The values of the order parameter $m$ and their comparison
with neutron scattering results, and the Fermi surfaces and their comparison with
photoemission experiments, lead us to the mean-field 
phase diagram shown in Fig.~\ref{PhaseDiagram3},
one of the main results of this effort. In this figure, the range of
$U$-$J/U$ couplings compatible
with neutron-ARPES experiments is labeled ``physical region''. This region is
relatively small, providing substantial constraints on the parameters to be
used in three-orbital Hubbard model investigations. If $U$ is smaller than in the
physical region, then the
state is not magnetic; if $U$ is larger, the state is insulating or it has a much
distorted Fermi surface. If $J/U$ is smaller than in the physical region, there is
no room for the small and intermediate value order parameters found in neutron scattering;
if $J/U$ is larger, then $U'$ becomes too small or negative and thus unphysical.

It is important to clarify that our mean-field approximation does not incorporate
the effect of fluctuations. For this reason this type of approximations are
somewhat ``rigid'' and it can be expected,\cite{0910.2707} although actual calculations are very
difficult, that the true ``physical region'' may be larger than shown 
in Fig.~\ref{PhaseDiagram3}. Thus, readers should consider the location of our
 ``physical regions'' as the center of a potentially broader area where agreement with experiments
can be found. But this does not invalidate our main point: intermediate $U$'s and
intermediate $J/U$'s are needed for agreement with available neutrons, transport,
and photoemission experiments.

A final remark is with regards to the actual value of $U$ of order just 1 eV 
in the physical region: this cannot be the bare $U$ but must already incorporate
the influence of screening in a model where the long-range Coulomb interactions
are included. It is for this reason that standard metals in general tend to have
very small $U$s when studied via Hubbard-like models, while it is known that 
the bare atomic values for $U$ are always of several eVs.\cite{other-results}

\section{Results for Five-Orbital Models}
\label{five}

In this section, results for two five-orbital models  
are presented, with a similar
organization as for three orbitals. The hopping
amplitudes and on-site energies of  a novel ``Model 1'' are provided in the Appendix
 (using as criterion for their determination finding qualitative agreement with band calculations, 
as shown in Fig.~6(a)).
The more accurate ``Model 2'' is the model introduced in Ref.~\onlinecite{Graser08},
where the reader can find the actual values of the parameters. These two models generate
similar Fermi surfaces, but the values for the hopping amplitudes are rather different.
Thus, they are useful to test whether our conclusions do or do not depend on small details.
Indeed, an overall conclusion
of our study is that the ``physical region'' is qualitatively similar for all
the models analyzed in this manuscript.
Both
models studied in this section are 
at electronic density $n=6.0$ in order to address the parent
compounds.\cite{detail} The $\delta$-function broadening used
here is 0.01 eV.

\subsection{Comparison with neutron scattering for Model 1}

\begin{figure}[ht]
\begin{center}
\includegraphics[clip,width=80mm,angle=0]{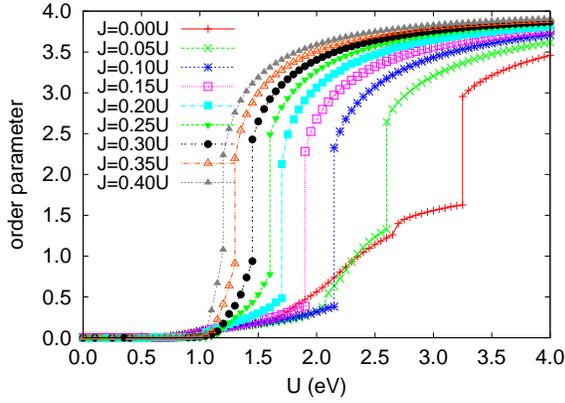}
\caption{Mean-field order parameter at wavevector $(\pi,0)$ 
vs. $U$ (in eV units) for the five-orbital Model 1 discussed
in the text, parametric with the values of $J/U$ indicated. Note
the presence of a critical $U$ where the order parameter first
develops, and a second critical $U$ where a discontinuity occurs
(see also Ref.~\onlinecite{Yuetal08}).} \label{mU5}
\end{center}
\end{figure}

The mean-field order parameter at wavevector $(\pi,0)$ vs. $U$ for Model 1 is 
shown in Fig.~\ref{mU5}, parametric with $J/U$. These results are
in several aspects qualitatively similar to those discussed before
for three orbitals in Fig.~\ref{mU3}, but there are some differences.
For instance, in this case a discontinuity in $m$ is found for
all the values of $J/U$ investigated. This shows that for this model
and using mean-field techniques, there is a range of values of $m$ 
(roughly between 1 and 2.5 depending on $J/U$) for which there are no
solutions. It turns out that neutron scattering experiments for 1111 and
122 materials have unveiled
low to intermediate values of $m$, and such order parameters fit 
in the range where the  mean-field analysis provides
stable solutions. Thus, for the case of five orbitals, the comparison
between the order parameter $m$ and neutron scattering simply restricts 
$U$ to be between the first critical value, where $m$ develops,
and the second critical coupling where the discontinuity occurs. This
range is larger for small $J$ than for larger values.~\cite{calderon2}
However, it should be noted that the magnetization in each orbital is
parallel even for small $J$ and $U$, in contrast to the state with
antiparallel orbital magnetization found in
Ref.~\onlinecite{calderon2}, and in contrast to the model discussed in
Sec.~\ref{Sec:graser}. The small overall magnetic moment results here
from weak magnetization of the individual orbitals rather than
from their partial cancellation.

\subsection{Comparison with ARPES experiments for Model 1}

\begin{figure}[ht]
\begin{center}
\includegraphics[clip,width=50mm,angle=0]{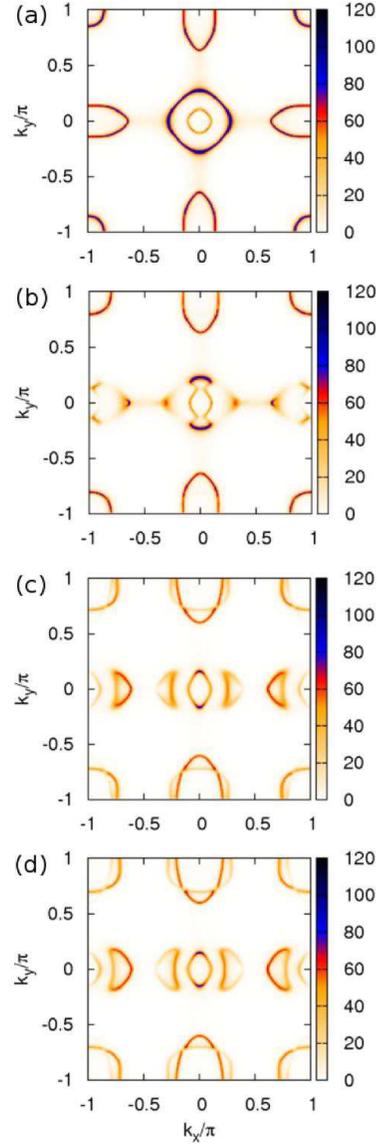}
\caption{Unfolded Fermi surfaces of the five-orbital Model 1 in the mean-field approximation
for the cases
(a) $U=J=0$, as reference; (b) $J/U=0.23$, $U=1.25$, $m=0.2$; (c) $J/U=0.28$, $U=1.45$, 
$m=0.7$; and (d) $J/U=0.30$, $U=1.4$, $m=0.7$. As in the case of three orbitals (Fig.~\ref{FS3})
and as in ARPES experiments, the results show distorted $\Gamma$-point hole pockets and satellite
features next to them (and also next to the electron pockets at $(\pi,0)$).}
\label{FS5}
\end{center}
\end{figure}

The comparison between the mean-field 
one-particle spectral function for the five-orbital models and ARPES experiments
introduces more severe restrictions on the values of $J/U$ and $U$ than those
discussed before in the neutron scattering context. Figure~\ref{FS5} shows the Fermi
surfaces of Model 1 for special values of $J/U$ and $U$ in a region compatible with ARPES. 
 Qualitatively, these results resemble those of the three-orbital model:
there are remnants of the original hole and electron pockets of the non-interacting
limit and, in addition, there are satellite features
near the $\Gamma$-point (and also near
the original non-interacting electron pocket at $(\pi,0)$).

\begin{figure}[ht]
\begin{center}
\includegraphics[clip,width=65mm,angle=0]{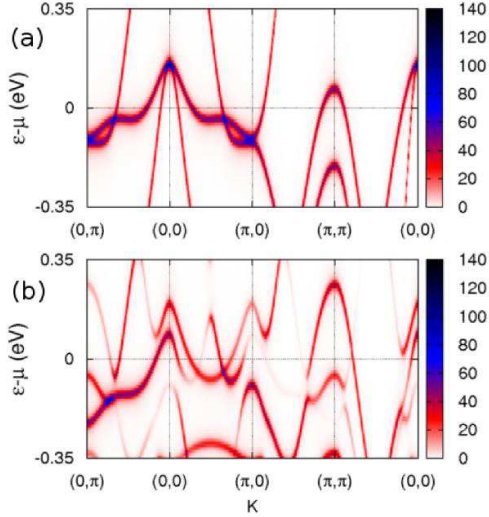}
\caption{Unfolded band structure mean-field results for the five-orbital Model 1 and cases 
(a) $U=J=0$, as reference, and  (b) $J/U=0.28$, $U=1.45$, $m=0.7$, 
illustrating in (b) the origin of the
satellite pockets shown in Fig.~\ref{FS5}.
The scale used (arbitrary units) is on the right.} 
\label{bands5}
\end{center}
\end{figure}

Figure~\ref{bands5} shows the actual bands for the five-orbital Model 1 for the
non-interacting case and one example of a set of couplings 
with satellite pockets
created by the magnetic order. While qualitatively similar to the results for three
orbitals (Fig.~\ref{bands3}) there are some interesting differences: here the
$V$-shaped features that originate the satellite pockets arise from a combination
of two bands while for three orbitals they emerge from the bending of a single band (see also Ref.~\onlinecite{newpaper}).
However, the level of accuracy of the ARPES experiments is not sufficient to
distinguish between these two cases, plus there is still room to further refine
the hopping amplitudes of the models used here to adjust for finer details of the
ARPES experiments.

\begin{figure}[ht]
\begin{center}
\includegraphics[clip,width=50mm,angle=0]{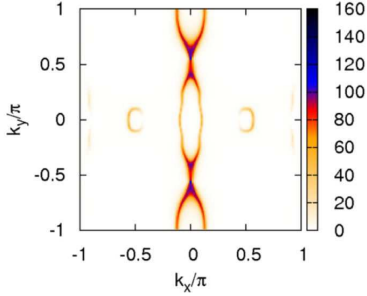}
\caption{Fermi surface of the five-orbital Model 1
in the mean-field approximation for the case
$J/U=0.10$, $U=2.1$, and $m=0.36$. The lack of qualitative similarity with
ARPES experiments shows that these couplings are not physically relevant
to describe the pnictides.} \label{bands31}
\end{center}
\end{figure}

Even in regions of parameter space where
the ground state is magnetic and metallic, 
and the order parameter is in the range of neutron's experiments,
the Fermi surface may still be qualitatively different from that observed in
ARPES. As an example, consider the case shown
in Fig.~\ref{bands31} corresponding to $J/U=0.10$. These mean-field 
results for $A({\bf k},\omega)$ are clearly different
from those shown before in Fig.~\ref{FS5}, and regions where 
this type of discrepancies
are found are removed from the ``physical region'' for the model.

\subsection{Summary phase diagram for the five orbitals Model 1}

\begin{figure}[ht]
\begin{center}
\includegraphics[clip,width=90mm,angle=0]{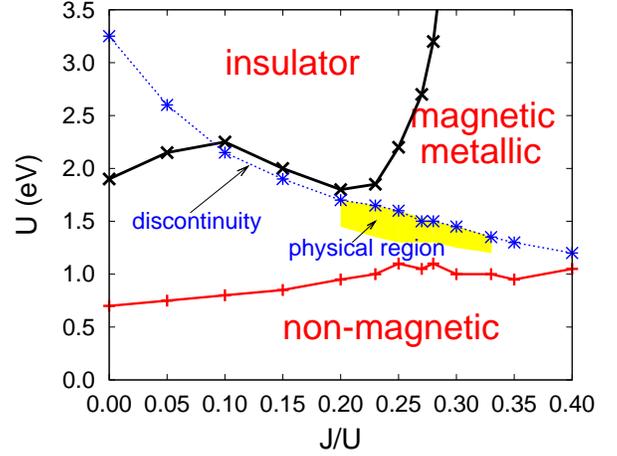}
\caption{Phase diagram for the five-orbital Model 1 obtained with the mean-field
approximation. As in Fig.~\ref{PhaseDiagram3} 
the ``physical region'' in yellow is the
regime of couplings found to be compatible with neutron and photoemission experiments.
The rest of the notation and details were already explained in Fig.~\ref{PhaseDiagram3}.
} \label{PhaseDiagram5}
\end{center}
\end{figure}

Similarly as for three orbitals, here a summary phase diagram is provided
for the five-orbital Model 1 in Fig.~\ref{PhaseDiagram5}. The labeling 
convention is the same as for
three orbitals in Fig.~\ref{PhaseDiagram3}. Using the information about 
neutron scattering and order parameters restricts $U$ and $J/U$ simply to
be between the non-magnetic region and the discontinuity line. 
From these perspectives alone the ``physical region'' would be much
larger than for the three orbitals case. However, considering the Fermi
surface shape and its comparison with ARPES introduces more severe constraints,
basically excluding the small $J/U$ regime below 0.15. As a consequence, the
final ``physical region'' ends up being similar to that obtained with
the three-orbital model. 
As explained in Sec.~III.C, note also that fluctuations beyond our mean-field approximation
are expected to render the ``physical region'' actually 
larger than displayed in Fig.~\ref{PhaseDiagram5}.

\subsection{Comparison with neutron scattering for Model 2}\label{Sec:graser}

\begin{figure}[ht]
\begin{center}
\includegraphics[clip,width=80mm,angle=0]{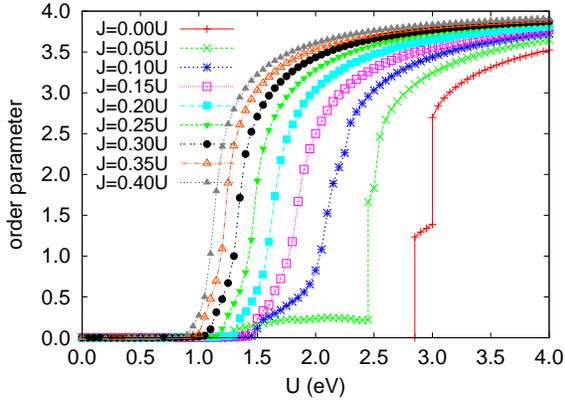}
\caption{Mean-field order parameter at wavevector $(\pi,0)$ 
vs. $U$ (in eV units) for the five-orbital Model 2, 
parametric with the values of $J/U$ indicated. Note that compared with Fig.~\ref{mU5},
the discontinuity in the order parameter now only occurs at small $J/U$.} 
\label{mU5-Graser}
\end{center}
\end{figure}

The mean-field order parameter at wavevector $(\pi,0)$ vs. $U$ is 
shown for Model 2 in Fig.~\ref{mU5-Graser}, parametric with $J/U$. Compared with 
the results for Model 1 in Fig.~\ref{mU5}, note that now the discontinuity in the order
parameter occurs only at small $J/U$. In this respect, the results for Model 1 are similar
to those recently reported for a similar model,\cite{calderon2} where small $J/U$ and
intermediate $U$ were emphasized (see also
Ref.~\onlinecite{Yuetal08}). This similarity persists in the
microscopic details of the state realized at small $J$: The
values for the magnetization of different orbitals can here have a
different sign, i.e., the intermediate-spin state with antiparallel
orbital magnetization discussed in Ref.~\onlinecite{calderon2} is
stabilized. While neutron scattering can only detect the overall moment and
is not expected to distinguish between this scenario and the one with
parallel magnetization found in other models, the precise
microscopic nature of the state realized in iron pnictide compounds
remains to be clarified. 

\subsection{Comparison with ARPES experiments for Model 2}

\begin{figure}[ht]
\begin{center}
\includegraphics[clip,width=55mm,angle=0]{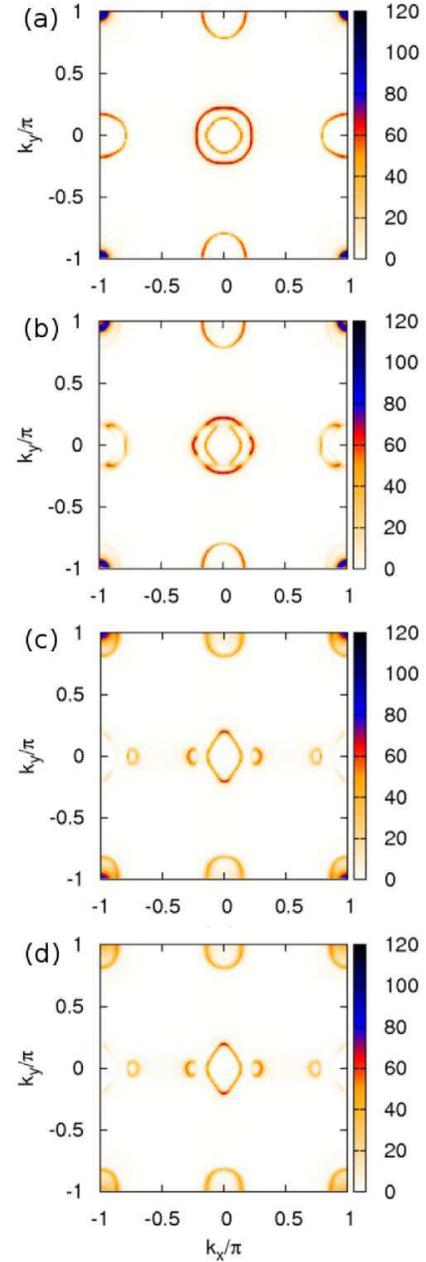}
\caption{Unfolded Fermi surfaces of the five-orbital Model 2 
in the mean-field approximation at $n=6.0$, for 
(a) $U=J=0$, as reference; (b) $J/U=0.20$, $U=1.35$, $m=0.2$; (c) $J/U=0.25$, $U=1.35$, 
$m=0.5$; and (d) $J/U=0.30$, $U=1.25$, $m=0.6$. As in the previous cases (Figs.~\ref{FS3} and \ref{FS5})
and as in ARPES experiments, the results show distorted $\Gamma$-point hole pockets and satellite
features next to them (and also next to the electron pockets at $(\pi,0)$).}
\label{FS5-Graser}
\end{center}
\end{figure}

\begin{figure}[ht]
\begin{center}
\includegraphics[clip,width=65mm,angle=0]{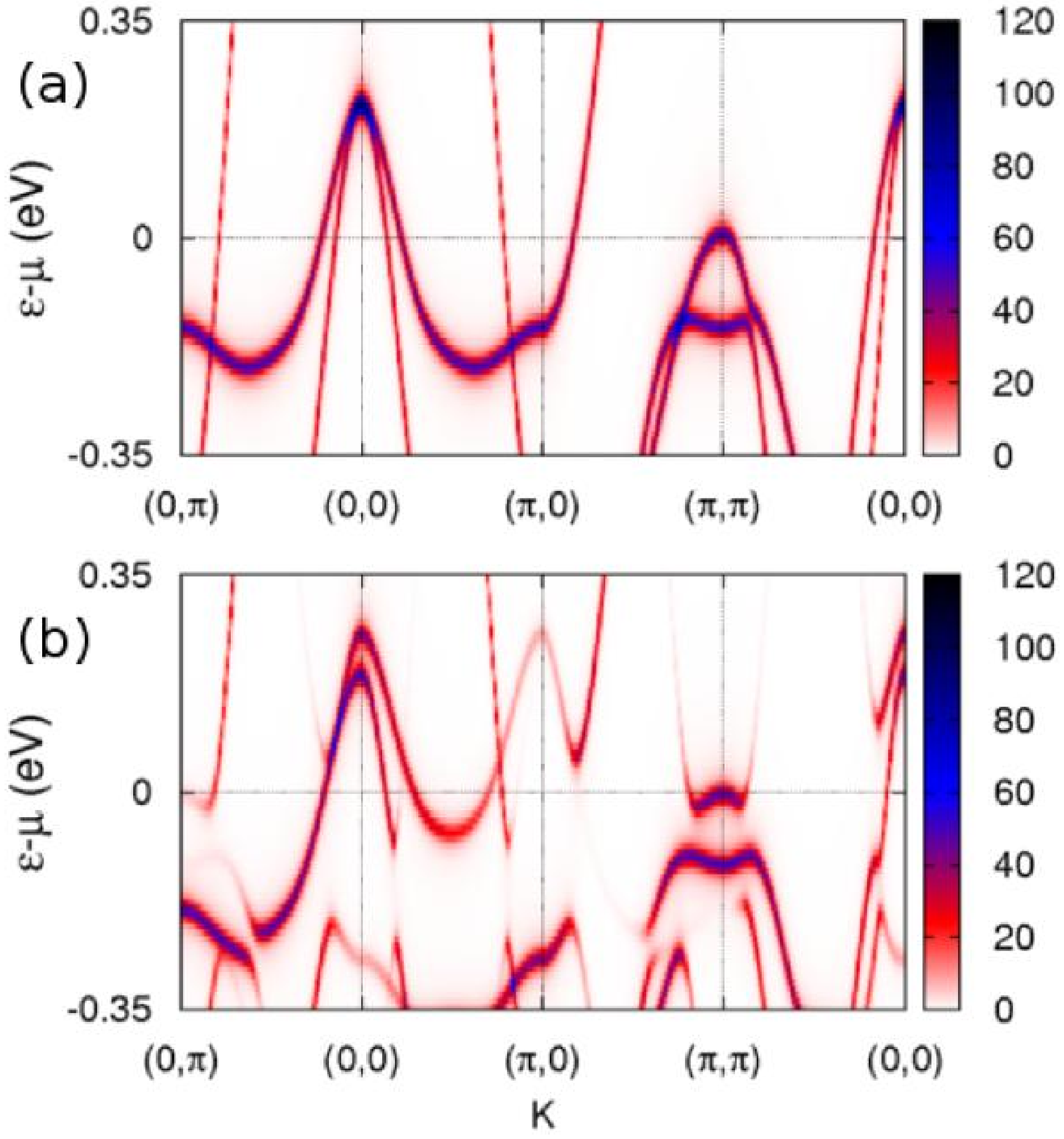}
\caption{Unfolded band structure mean-field results for the five-orbital Model 2 and cases 
(a) $U=J=0$, as reference, and  (b) $J/U=0.25$, $U=1.35$, $m=0.5$, 
illustrating in (b) the
satellite pockets shown in Fig.~\ref{FS5-Graser}.
The scale used (arbitrary units) is on the right. Note
the similarity of these results with those of Fig.~\ref{bands5}} 
\label{bands5-Graser}
\end{center}
\end{figure}

Similarly as presented for Model 1 and for the three-orbital model, 
Figure~\ref{FS5-Graser} shows the Fermi
surfaces for special values of $J/U$ and $U$ in a region found to be qualitatively 
compatible with ARPES. 
These Fermi surfaces show remnants of the non-interacting Fermi surfaces, plus satellite
extra features near the $\Gamma$- and $(\pi,0)$)-points. Note that the character of
these satellites, namely whether they are electron or hole pockets, 
depends on details and our focus has only been on the existence of extra features
in the correct location as compared with ARPES.
Figure~\ref{bands5-Graser} shows the bands for the five-orbital Model 2, both for 
the non-interacting case and for one example from the set of couplings used 
in  Fig.~\ref{FS5-Graser}. Most of the comments already made for the band structure 
of Model 1 with regards to the $V$-shaped features apply to Model 2 as well.
However, it is interesting to notice that the satellite pockets are mainly
electron-like for Model 1 and hole-like for Model 2. 

\begin{figure}[ht]
\begin{center}
\includegraphics[clip,width=90mm,angle=0]{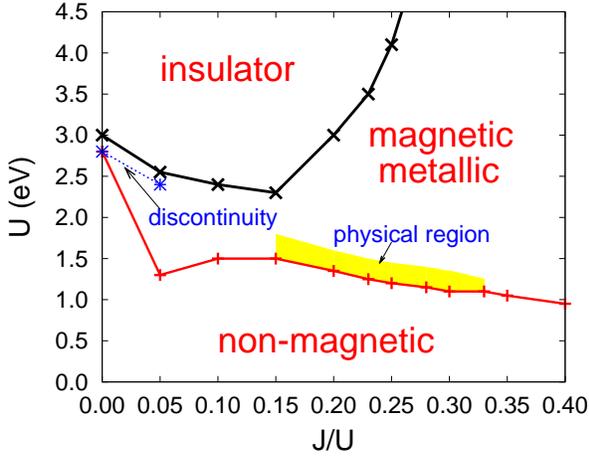}
\caption{Phase diagram for the five-orbital Model 2 
obtained with the mean-field
approximation. As in Figs.~\ref{PhaseDiagram3} and~\ref{PhaseDiagram5}, 
the ``physical region'' in yellow is the
regime of couplings found to be compatible with neutron and photoemission experiments.
The rest of the notation and details were already explained in Fig.~\ref{PhaseDiagram3}.
} 
\label{PhaseDiagram5-Graser}
\end{center}
\end{figure}

\subsection{Summary phase diagram for the five orbitals Model 2}

As for the other models considered in this effort, in Fig.~\ref{PhaseDiagram5-Graser} 
a summary phase diagram is provided
for the five-orbital Model 2. 
The labeling convention is the same as in Figs.~\ref{PhaseDiagram3} and \ref{PhaseDiagram5},
as well as the procedure to establish the so-called ``physical region''. 
The approximate location of this
region with regards to $U$ and $J/U$ is similar to that reported in Figs.~\ref{PhaseDiagram3} and 
~\ref{PhaseDiagram5}, showing again that our 
conclusions do not strongly depend on details. The  physical
region in Fig.~\ref{PhaseDiagram5-Graser} is narrow
along the $U$ axis direction because $m$ changes rapidly with increasing $U$,
at small $m$. 
But, as discussed in Sec.~III.C, fluctuations beyond mean-field approximations 
are expected to expand the size of the physical regions of the various models.

\section{RPA Pairing Symmetries Analysis}
\label{rpa}

After having established a regime of couplings $U$ and $J/U$ where the
mean-field approximation to the multiorbital Hubbard model gives a 
qualitative agreement with neutron and ARPES experiments, and also with the metallic nature of the
undoped compounds, it is interesting
to investigate what kind of pairing tendencies are observed in those
regions of parameter space. Unfortunately, the number of many-body tools available
to carry out such investigation is very small. While for the 
two-orbital model it is possible to carry out Lanczos studies on small clusters~\cite{Daghofer08}
to at least analyze the quantum numbers of the ground state with two extra particles,
for three orbitals or more this calculation is no longer practical. There are 
also no sign-problem-free Monte Carlo techniques available for these 
complex calculations. Thus, here the Random Phase Approximation will be used.
Experience with the cuprates indicates that this method did capture the relevance
of $d$-wave pairing in that context and for this reason it will be used here
as well, although with the caveat that the method is qualitative at best. 
RPA has been recently applied to the five-orbital Model 2 considered here~\cite{Graser08} and the
dominance of $A_{ 1g}$ states was unveiled at small $J/U$, with $B_{ 1g}$ 
a close competitor. In our investigations shown below, 
 in the ``realistic'' regime of couplings 
{\it the RPA dominant pairing tendency always has nodes or quasi-nodes}, 
an interesting result in view
of the current experimental controversy between ARPES and other techniques
with regards to the nodal structure of the superconducting state. 
It is also important to remark that other irreducible representations are
closely competing with those that dominate, and small changes in parameters alter
their relative dominance.

\subsection{Random Phase Approximation Formalism}

\begin{figure}[ht]
\begin{center}
\includegraphics[clip,width=60mm,angle=270]{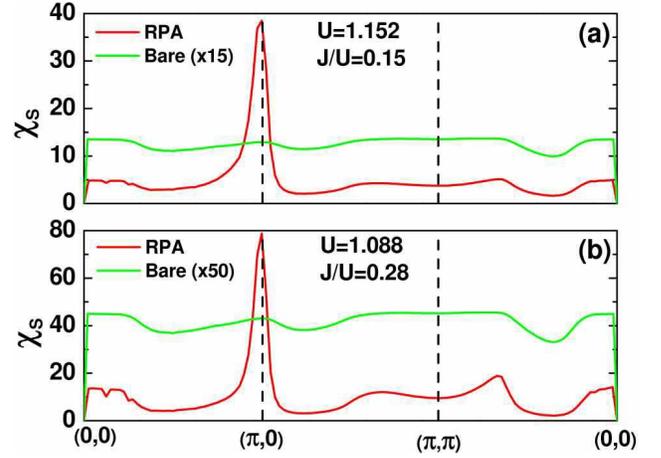}
\caption{Bare and RPA magnetic susceptibilities vs. BZ 
wavevector, at the values of
$U$ and $J/U$ indicated, for the five-orbital Model 1.} \label{fig1-5orb}
\end{center}
\end{figure}

For completeness, here the RPA approximation will be briefly reviewed, 
with emphasis on some technical aspects. 
This section follows the formalism already described in Refs.~\onlinecite{Graser08} and \onlinecite{aoki:2009}.
Assuming that spin fluctuations (excitations in the paramagnetic state above 
the critical temperature, also called paramagnons)
are responsible for the pairing mechanism present
in the iron-pnictides, 
the RPA method will be used to study these fluctuations
(caused by itinerant carriers) beyond the mean-field level. 
In linear response theory, prior to
including the many-body interactions, these spin-wave-like excitations 
are obtained
from the non-interacting Lindhard function (here defined 
for a multi-orbital model)
\begin{equation}
\chi^0_{l_1,l_2,l_3,l_4}({\bf q}) =\sum_{\bf k} G_{l_1,l_3}({\bf k}+{\bf q})G_{l_4,l_2}({\bf k}),
\end{equation}
where each of the four indices needed to characterize the Lindhard
function takes the values 
$l_i = 1,...,n_o$ $(i = 1,...,4)$, with $n_o$ the total number of orbitals
being considered (here, $n_o=3$ or $n_o=5$).
The non-interacting Green's function, which describes the propagation
of the elementary excitations 
present in the non-interacting model, can be written as
\begin{equation}
G_{l_1,l_3} ({\bf k},\omega) = \sum_\mu \frac{\langle l_1 | \mu {\bf k} \rangle \langle  \mu {\bf k} | l_3  \rangle}{\omega - E_\mu({\bf k})},
\end{equation}
where $b^{l_1}_{\mu}({\bf k})=\langle l_1 | \mu {\bf k} \rangle$ projects the tight-binding state $| \mu {\bf k} \rangle$ into the orbital $| l_1 \rangle$,
where $\mu$ labels which one of the $2n_o$ tight-binding bands (in the extended BZ, with one iron per unit cell) the tight-binding
state belongs to (with energy $ E_\mu({\bf k})$).
Introducing many-body interactions in the model, the Lindhard function gives origin
to (distinct) spin and charge susceptibilities, which, calculated at the RPA level through a
Dyson equation, can be expressed as (using matrix equations),
\begin{equation}
\hat{\chi}_s({\bf q})=\frac{\hat{\chi}^0({\bf q})}{1-\hat{U}^s\hat{\chi}^0({\bf q})} ,
\end{equation}
\begin{equation}
\hat{\chi}_c({\bf q})=\frac{\hat{\chi}^0({\bf q})}{1+\hat{U}^c\hat{\chi}^0({\bf q})} ,
\end{equation}
where expressions for the non-zero matrix elements of $\hat{U}^s$ and $\hat{U}^c$, in terms
of the many-body interactions, can be found in Refs.~\onlinecite{Graser08} and \onlinecite{aoki:2009}.
The magnetic susceptibility (both RPA and bare) to be shown below in, e.g., Fig.~\ref{fig1-5orb} is given by
\begin{equation}
{\chi}_s({\bf q})=\frac{1}{2} \sum_{l_1,l_2} (\hat{\chi}_s)({\bf q})_{l_1,l_1}^{l_2,l_2}.
\end{equation}

These RPA susceptibilities are used to construct a 
spin-singlet pairing interaction
describing the exchange of charge (orbital) and spin fluctuations, resulting
in an effective electron-electron interaction:\cite{berk-schrieffer}
\begin{equation}
\hat{V}^s({\bf q})=\frac{3}{2}\hat{U}^s\hat{\chi}_s({\bf q})\hat{U}^s
-\frac{1}{2}\hat{U}^c\hat{\chi}_c({\bf q})\hat{U}^c+\frac{1}{2}(\hat{U}^s+\hat{U}^c).
\end{equation}

Assuming that the dominant scattering occurs close to the Fermi surface,
one can calculate the scattering amplitude of a Cooper pair between
two points {\it at} the Fermi surface $\left[({\bf k},-{\bf k}) \rightarrow ({\bf k}^{\prime},-{\bf k}^{\prime})\right]$,
where the momenta $\bf k$ and ${\bf k}^{\prime}$ are restricted to the Fermi surface pockets $i$ and $j$,
which span the existing 
pockets for the chemical potential used. To indicate this restriction
a subscript $i$ (or $j$) is added to the bands $\mu$ and $\nu$ in question, and the (symmetrized) 
interaction vertex
becomes:
\begin{eqnarray}
{\Gamma}_{ij} ({\bf k},{\bf k}') & = & \sum_{l_1,l_2,l_3,l_4} b^{l_2,*}_{\mu_i}(-{\bf k})~b^{l_1,*}_{\mu_i}({\bf k})
\mathrm{Re}\left[V_{l_1,l_2}^{l_3,l_4} ({\bf k},{\bf k}')\right]\nonumber\\
&\times& b^{l_3}_{\nu_j}({\bf k}^{\prime})~b^{l_4}_{\nu_j}(-{\bf k}^{\prime}),
\end{eqnarray}
which, after plugging into the linearized Eliashberg equation, results in 
a dimensionless pairing strength functional:\cite{scalapino-hirsch:86,aguilar:1994}
\begin{equation}
\lambda [g({\bf k})] = - \frac{\sum_{ij} \oint_{C_i} \frac{d k_\parallel}{v_F({\bf k})} \oint_{C_j}
\frac{d k_\parallel'}{v_F({\bf k}')} g({\bf k}) {\Gamma}_{ij} ({\bf k},{\bf k}')
g({\bf k}')}{ 4\pi^2 \sum_i \oint_{C_i} \frac{d k_\parallel}{v_F({\bf k})} [g({\bf k})]^2 },
\end{equation}
where the superconducting gap $\Delta({\bf k})$ (in reciprocal space) was separated
into an amplitude $\Delta$ and a function $g({\bf k})$, which should have the symmetry
of one of the irreducible representations of the corresponding point group (in the case
of the pnictides, 
the $D_{ 4h}$ group is considered, with representations $A_{1g}$, $A_{2g}$,
$B_{1g}$, $B_{2g}$, and $E_{g}$). The stationary condition leads to an eigenvalue problem
defined over the Fermi surface:
\begin{equation}
- \sum_j \oint_{C_j} d k_\parallel'\frac{1}{4\pi^2 v_F({\bf k}')} {\Gamma}_{ij} ({\bf k},{\bf k}')
g_j({\bf k}') = \lambda g_i({\bf k}),
\end{equation}
which is an eigenvalue problem involving a matrix $\left[\Gamma({\bf k},{\bf k}^{\prime})\right]$,
and where $g_i({\bf k})$ is the value of the gap function for a point $\bf k$ {\it at} the pocket $i$,
and $\lambda$ is the associated eigenvalue. The highest eigenvalue
(normalized to 1) indicates what gap function will have the
highest critical temperature. The dimension of the matrices to be diagonalized
will be determined by the number of ${\bf k}$ points taken along the Fermi
surface.
For the 
calculations shown here, it was observed that considering approximately 200 points along the Fermi
surface provides a good convergence. All calculations were done at temperature $T=0.02$eV,
and an imaginary part $\eta=10^{-5}$ was used to regularize the Green's functions.
All the sums over the BZ were done with uniform $64 \times 64$ meshes
(calculations with a $128 \times 128$ mesh yielded qualitatively the same results).

\subsection{RPA Pairing Symmetries for the Five-Orbital Model 1}

The RPA analysis of the ``realistic'' $U$ and $J/U$ regime will start
here with the five-orbital Model 1. In Fig.~\ref{fig1-5orb}, the RPA results
for the magnetic susceptibility $\chi_S$ are shown for two values of $J/U$ (and
slightly different $U$). The case $J/U=0.15$ is close but outside the
``physical region'' of Fig.~\ref{PhaseDiagram5}, while $J/U=0.28$ is clearly 
inside that region. However, with regards to $\chi_S$, the figure shows that there
are no substantial changes in the magnetic response for these two $J/U$s:
an approximately flat bare $\chi_S$ becomes a RPA $\chi_S$ with a sharp 
peak at the correct wavevector $(\pi,0)$. 

\begin{figure}[ht]
\begin{center}
\includegraphics[clip,width=65mm,angle=270]{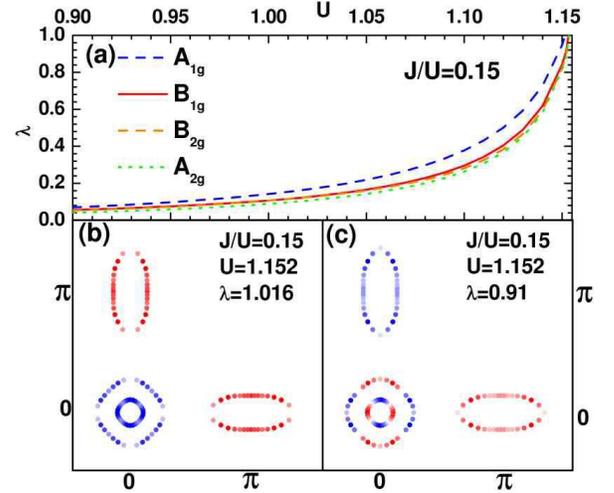}
\caption{
(a) RPA pairing eigenvalues vs. $U$, at $J/U=0.15$, for the five-orbital Model 1. The symmetries
of each eigenvalue are indicated. (b) Dominant $A_{ 1g}$ 
gap function with a similar color convention as in Ref.~\onlinecite{Graser08} (blue and red denoting different signs). (c) Subdominant state belonging to the $B_{ 1g}$ representation. } \label{fig2-5orb}
\end{center}
\end{figure}

The RPA pairing eigenvalues $\lambda$ are shown in Fig.~\ref{fig2-5orb}~(a) as a function of
$U$, at a constant $J/U=0.15$. Shown in (a) are the four first eigenvalues, that
happened to have the four different symmetries that are indicated (as opposed
to repeating some symmetries). The dominant
one is $A_{ 1g}$, in agreement with Ref.~\onlinecite{Graser08} 
using the five-orbital Model 2.
The actual pairing function for this dominant eigenvalue is shown
in Fig.~\ref{fig2-5orb}~(b). This is qualitatively 
of the form of the well-known $s^{+-}$ pairing, with different
signs between the hole and electron pockets. However, 
the actual gap values are
not uniform along each pocket. Actually, all the
pockets present some narrow regions where the gap functions nearly vanish, a feature
quite reminiscent of nodes (dubbed here quasi-nodes\cite{quasi-nodes}). 
Overall, these results, and those presented below for Model 2 and for the three-orbital model,
are compatible with the recent observations in Refs.~\onlinecite{aoki:2009,Graser10}
that pockets at $(\pi,\pi)$ suppress nodes in the $A_{1g}$ sector while the absence of those pockets
produce a nodal $A_{1g}$ (note that in some of our results here and 
below, the superconducting order parameter at $(\pi,\pi)$ is weak and difficult to see).

In Fig.~\ref{fig2-5orb}~(c), the pairing function of the 
subdominant eigenvalue is shown. By mere inspection it is clear that it belongs
to the $B_{ 1g}$ sector, in qualitative agreement with  
Ref.~\onlinecite{Graser08} that also identified this symmetry as the
first competitor to $A_{ 1g}$. Thus, for $J/U=0.15$ our results are
similar to those previously reported~\cite{Graser08} and have the value
of providing a test of our procedure. However, Fig.~\ref{fig2-5orb}~(a) contains 
extra information not discussed before: it also
shows that two other eigenvalues are nearly degenerate with $B_{ 1g}$ 
and in view of the approximate nature of the calculation 
they should also be considered
as important competitors. These extra competitors belong to 
the $B_{ 2g}$ and $A_{ 2g}$ sectors. In particular the $B_{ 2g}$
is compatible with the Lanczos results found for the two-orbital model
in a similar ``intermediate $U$'' region of parameter space.~\cite{Daghofer08}

\begin{figure}[ht]
\begin{center}
\includegraphics[clip,width=65mm,angle=270]{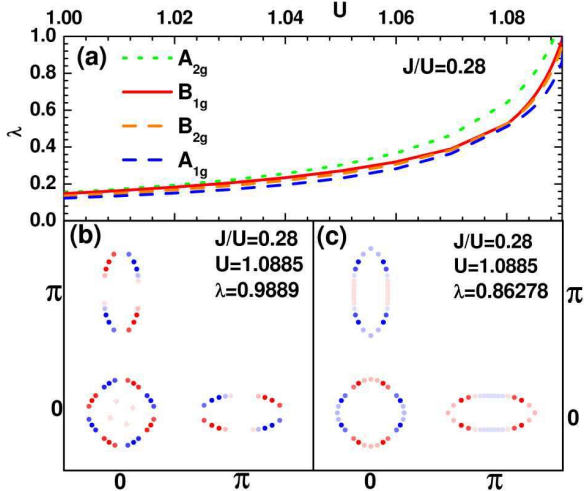}
\caption{
(a) RPA pairing eigenvalues vs. $U$, at $J/U=0.28$, for the five-orbital Model 1. The symmetries
of each eigenvalue are indicated. (b) Dominant $A_{ 2g}$ 
gap function with a similar color convention as in Ref.~\onlinecite{Graser08} (blue and red denoting different signs).
(c) Subdominant state belonging to the $B_{ 1g}$ representation.
} \label{fig3-5orb}
\end{center}
\end{figure}

Figure~\ref{fig3-5orb} provides novel interesting information. Switching now to $J/U=0.28$, i.e. inside
the $U$-$J/U$ region considered realistic based on neutron and ARPES experiments,
the dominant eigenvalue now belongs to the $A_{ 2g}$ irreducible representation (with a gap function shown in (b)).
$B_{1g}$ (see (c)) is still the subdominant tendency, but it is nearly degenerate with $B_{2g}$.
For this value of $J/U$ the $A_{ 1g}$ channel is 4th in the relative order. 
Thus, a relatively small variation 
in $J/U$ induces a qualitatively drastic
rearrangement of pairing eigenvalues. However, all of them have nodes (or quasi-nodes as
in the case of $A_{1g}$). Thus, it appears unavoidable to conclude that nodal superconductivity
should be present in the pnictides, at least within pairing tendencies that rely on electronic
mechanisms (and within the RPA and mean-field approximations).

The presence of pairing states with nodes also extends to the regime of electron doping. For
instance, for density $n = 6.125$ our RPA studies (not shown) indicate the dominance of a $B_{ 1g}$ state
for $J/U=0.28$ and $0.35$, with an $A_{ 1g}$ state with nodes being the subdominant pairing. While
small changes alter the relative balance between the many competing states, 
all of those states, 
including the $A_{ 1g}$, present a nodal (or quasi-nodal) structure at the RPA level used here.

It is worth discussing these results in a historical context. In previous RPA studies of one-orbital Hubbard models in the spin-singlet channel,
it was observed a dominance of the well-known $d_{x^2-y^2}$-wave state
over other channels.\cite{scalapino-hirsch:86} Calculations in the one-orbital 
context did not report the presence
of other pairing states so close to the dominant one as shown here and in Ref.~\onlinecite{Graser08}.
In fact, the only competitors to the $d$-wave state that have been identified for the case
of one orbital are
spin-triplet $s$-wave odd-frequency states (not studied in our present investigations due
to their spin-triplet nature). The appearance of odd-frequency states as competitors was analyzed in studies
on square lattices,\cite{bulut} as well as on triangular lattices.\cite{vojta}
But among the spin-singlet states, the dominance of $d$-wave was clear for one orbital.
Thus, finding in the multiorbital models so many spin-singlet even-frequency states 
competing with the dominant ones is surprising and merits further work.
These results suggest that within the pnictide family different pairing
channels may be stable in different pnictide 
compounds since small changes in parameters, such
as caused by chemical doping or pressure, may lead
to changes in the dominant pairing tendency.

\begin{figure}[ht]
\begin{center}
\includegraphics[clip,width=30mm,angle=270]{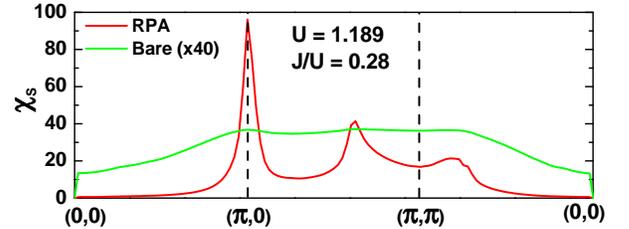}
\caption{Bare and RPA magnetic susceptibilities vs. BZ 
wavevector, at the values of
$U$ and $J/U$ indicated, for the five-orbital Model 2.} \label{fig1-5orb-Graser}
\end{center}
\end{figure}

\subsection{RPA Pairing Symmetries for the Five-Orbital Model 2}

For completeness, the RPA analysis for Model 2 is here included. 
Model 2 was already studied in Ref.~\onlinecite{Graser08}, but here the
focus will be on the ``realistic'' regime of $U$ and $J/U$.
In Fig.~\ref{fig1-5orb-Graser}, the RPA results
for the magnetic susceptibility $\chi_S$ are shown for representative values
of $U$ and $J/U$. In this regime, there is a clearly dominant
peak at wavevector $(\pi,0)$, although there are subdominant peaks at other momenta as well.
The results at other value of $U$ and $J/U$ in the vicinity of the one shown are
similar.

\begin{figure}[ht]
\begin{center}
\includegraphics[clip,width=65mm,angle=270]{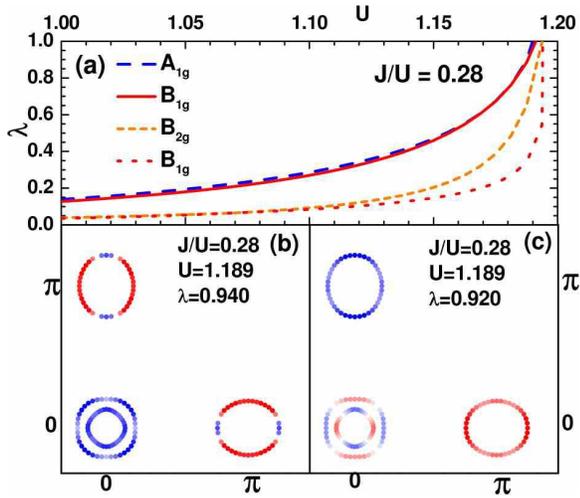}
\caption{
(a) RPA pairing eigenvalues vs. $U$, at $J/U=0.28$, for the five-orbital Model 2. The symmetries
of each eigenvalue are indicated. (b) Dominant $A_{ 1g}$ 
gap function with a similar color convention as in Ref.~\onlinecite{Graser08} (blue and red denoting different signs). (c) Subdominant state belonging to the $B_{ 1g}$ representation. } 
\label{fig2-5orb-Graser}
\end{center}
\end{figure}

The corresponding RPA pairing eigenvalues $\lambda$ are shown in Fig.~\ref{fig2-5orb-Graser}~(a) 
as a function of
$U$, at a constant $J/U=0.28$. Following a similar organization as in the case of Model 1, 
shown in (a) are the four first eigenvalues. The dominant
one is $A_{ 1g}$ with nodes, in agreement with Refs.~\onlinecite{Graser08,Graser10}.
The pairing function for this dominant eigenvalue is shown
in Fig.~\ref{fig2-5orb-Graser}~(b): it has different
signs between the hole and electron pockets, and
the actual gap values are
not uniform along each pocket, actually presenting nodes.
In Fig.~\ref{fig2-5orb-Graser}~(c), the pairing function of the 
subdominant eigenvalue is shown. It belongs 
to the $B_{ 1g}$ sector, as found in
Ref.~\onlinecite{Graser08}. The importance of our results is that using exactly the same formalism our approach
does properly reproduce previous results with regards to 
the dominance of the $A_{ 1g}$ sector for Model 2, while
Model 1 (with a very similar Fermi surface) has pairing in other channels. This highlights again 
that very small
changes in parameters can dramatically alter the pairing instability channel.

\begin{figure}[ht]
\begin{center}
\includegraphics[clip,width=60mm,angle=270]{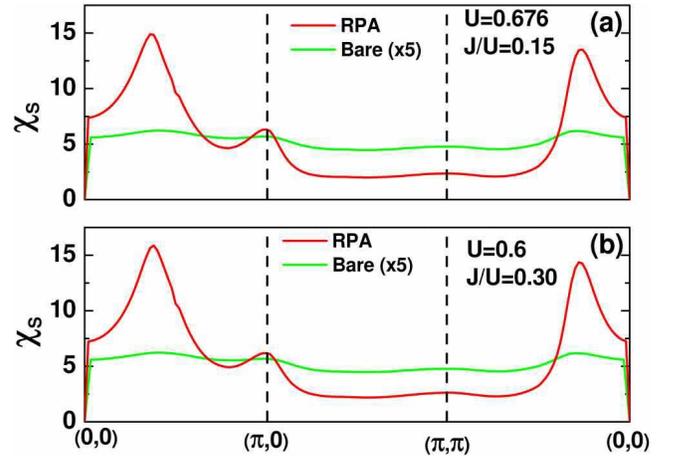}
\caption{Bare and RPA magnetic susceptibility vs. BZ 
wavevector, at the values of
$U$ and $J/U$ indicated, for the three-orbital model. Note that 
in addition to a small 
peak at $(\pi,0)$, there are larger peaks at other wavevectors.} \label{fig1-3orb}
\end{center}
\end{figure}

\subsection{RPA Pairing Symmetries for the Three-Orbital Model}

To complete the RPA analysis, now the three-orbital model will be considered. 
The results reported below will  not be as clear as in the previous case of five orbitals, but these
results are presented here anyway to alert the reader of the 
subtleties associated
with RPA approximations. In spite of the difficulties to be shown below,
the pairing states that dominate still have nodes. Thus, the general conclusion that
nodal superconductivity tends to be favored in Hubbard models remains the same, at least within the approximations used in our effort.

\begin{figure}[ht]
\begin{center}
\includegraphics[clip,width=65mm,angle=270]{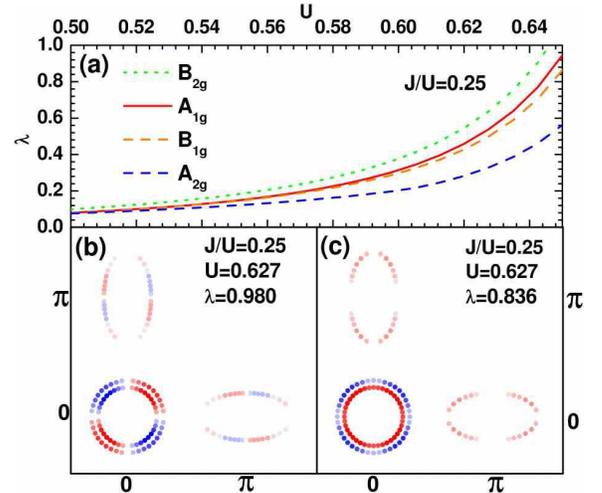}
\caption{
(a) RPA pairing eigenvalues vs. $U$, at $J/U=0.25$, for the three-orbital model. The symmetries
of each eigenvalue are indicated. (b) Dominant $B_{ 2g}$ 
gap function with a similar color convention as in Ref.~\onlinecite{Graser08} (blue and red denoting different signs).
(c) Subdominant state belonging to the $A_{ 1g}$ representation.
} \label{fig2-3orb}
\end{center}
\end{figure}

Figure~\ref{fig1-3orb} shows the RPA magnetic susceptibility for two cases of interest, $J/U = 0.15$ and $0.30$, within
the ``physical region'' of Fig.~\ref{PhaseDiagram3}. These results already illustrate the main problem found here:
although there is a small peak at wavevector $(\pi,0)$, there are other larger peaks at wavevectors closer to $(0,0)$.
Thus, the results to be shown for the pairing tendencies are in 
a magnetic background that does not correspond to
that of the pnictides experiments, but it contains states  with a variety of wavevectors. 

Figure~\ref{fig2-3orb}~(a) displays the $U$ dependence of the pairing
eigenvalues at $J/U=0.25$, inside the  ``physical region'' for three orbitals. Note that in this case it is
the $B_{ 2g}$ symmetry that dominates, as in Ref.~\onlinecite{Daghofer08}, with
$A_{ 1g}$ being subdominant. 
 The order parameter of the dominant $B_{ 2g}$ tendency
is in Fig.~\ref{fig2-3orb}~(b), while the subdominant $A_{ 1g}$ is in Fig.~\ref{fig2-3orb}~(c) 
(with the exotic detail that in this case the two hole pockets carry
a different sign for the order parameter). 
The presence of the $B_{ 2g}$ tendency was not only observed at the couplings of
Fig.~\ref{fig2-3orb} and $n=4$, but also in a wide range of $J/U$ and varying the electronic
density $n$ to 4.1 and 3.9 (not shown). Thus, such nodal pairing tendencies are robust. However,
a more detailed analysis of the RPA response of the three-orbital model, motivated by
the multiple wavevectors in its magnetic state, should be carried out in the future.

\section{Conclusions}
\label{conclusions}

In this publication, 
the undoped limit of the multiorbital Hubbard model for pnictides has been studied using a
standard mean-field approximation, similar to that employed in the study of
the single-orbital Hubbard model for the undoped cuprates. Within this approximation,
the magnitude of the order parameter associated with the $(\pi,0)$ magnetic order was studied varying $U$ and $J/U$.
In addition, the one-particle spectral function $A({\bf k},\omega)$ was also analyzed. Comparing results against
neutron scattering and ARPES experiments, allow us to define regions in parameter space, dubbed ``physical regions'',
where the mean-field model Hamiltonian predictions are in qualitative agreement with the above mentioned
experiments. These regions are relatively small in size since the ground state in this regime must be simultaneously metallic, 
magnetic with order parameters in the range found by neutrons, and with Fermi surfaces containing satellite pockets induced by the magnetic state near
the $\Gamma$-point hole pockets of the original 
band structure. 
Although fluctuations beyond the mean-field approximation are expected to enhance
the physical regions, our results still provide important constraints on 
the couplings to be used for
theoretical studies of multiorbital Hubbard models for pnictides. 

In addition, in the regime of $U$ and $J/U$ described above, the RPA approximation allowed us to make
predictions about the dominant pairing tendencies in the ``physical regions''. While it is clear
that several channels are competing in these models, namely that small changes in parameters can lead to
drastic changes in the dominant symmetries of the pairing states, the common property that emerges is the
presence of nodal superconducting states in those physical regions. Within these models it appears very difficult
to stabilize states without nodes. Even in regimes where the $A_{ 1g}$ state dominates, it still has quasi-nodes (very small
values of the amplitude at particular Fermi surface points) or true nodes. Thus, in these regards our results are more
compatible with the bulk measurements that reported nodal superconductivity 
than with the ARPES experiments reporting nodeless
superconductivity in doped pnictides. 
However, our studies are based on approximations that need to be refined. 
Hopefully, our study will 
initiate a debate on what are the true dominant tendencies in the 
multiorbital Hubbard model for pnictides, helping to decide
if the mechanism is electronic or phononic.

\section*{Acknowledgments}

This research was sponsored by 
the U.S. Department of Energy, 
Office of Basic Energy Sciences, 
Materials Sciences and Engineering Division (A.M. and E.D.), 
the Deutsche Forschungsgemeinschaft 
under the Emmy-Noether program (M.D.), the Sun Yat-Sen University under
the Hundred Talents program (D-X.Y.), and by the W. M. Keck Foundation (R.Y.). 
G.B.M. especially acknowledges the help 
of S. Graser in developing the RPA code, as well as 
discussions with A. Liebsch. E.D. acknowledges useful
discussions with D. Scalapino.

\section*{Appendix: parameters of five-orbital Model 1}

Using a similar notation as for three orbitals, and the parameters of Table II below,
the tight-binding portion of the
five-orbital Model 1 is
$H_{\rm TB}(\mathbf{ k}) = \sum_{\mathbf{ k},\sigma,\mu,\nu} T_{\mu,\nu}
(\mathbf{ k})
d^\dagger_{\mathbf{ k},\mu,\sigma} d^{\phantom{\dagger}}_{\mathbf{ k},\nu,\sigma}~,$
where
\begin{eqnarray}
T_{11/22} &=& 2t^{11}_{x/y}\cos  k_x +2t^{11}_{y/x}\cos  k_y +4t^{11}_{xy} \cos  k_x\cos  k_y \nonumber\\
  & &\pm 2t^{11}_{xx} (\cos 2 k_x - \cos 2 k_y) + 4t^{11}_{xxy/xyy} \cos 2 k_x \cos  k_y \nonumber\\
  & &+ 4t^{11}_{xyy/xxy} \cos  k_x \cos 2 k_y + 4t^{11}_{xxyy} \cos 2 k_x \cos 2 k_y \nonumber\\
  & &+ \epsilon_{11/22}, \label{eq:T11}\\
T_{33} &=& 2t^{33}_x(\cos  k_x+\cos  k_y) + 4t^{33}_{xy}\cos  k_x\cos  k_y \nonumber\\
       & & + 2t^{33}_{xx}(\cos 2 k_x+\cos 2 k_y) + \epsilon_{33}, \label{eq:T33} \\
T_{44} &=& 2t^{44}_x(\cos  k_x+\cos  k_y) + 4t^{44}_{xy}\cos  k_x\cos  k_y \nonumber\\
       & & + 2t^{44}_{xx}(\cos 2 k_x+\cos 2 k_y) \nonumber\\
      & &+ 4t^{44}_{xxy}(\cos 2 k_x \cos  k_y +\cos k_x\cos 2 k_y) \nonumber\\
      & &+ 4t^{44}_{xxyy} \cos 2 k_x \cos 2 k_y + \epsilon_{44}, \label{eq:T44} \\
T_{55} &=& 2t^{55}_x(\cos  k_x+\cos  k_y) + 2t^{55}_{xx}(\cos 2 k_x+\cos 2 k_y) \nonumber\\
       & & + 4t^{55}_{xxy}(\cos 2 k_x \cos  k_y +\cos k_x \cos 2 k_y) \nonumber\\
      & & + 4t^{55}_{xxyy} \cos 2 k_x \cos 2 k_y + \epsilon_{55}, \label{eq:T55} \\
T_{12} &=& T_{21} \nonumber\\
      &=& -4t^{12}_{xy}\sin  k_x \sin  k_y \nonumber\\
      & & - 4t^{12}_{xxy}(\sin 2 k_x \sin  k_y + \sin k_x \sin 2 k_y) \nonumber\\
      & & -4t^{12}_{xxyy}\sin 2 k_x \sin 2 k_y , \label{eq:T12}\\
T_{13/23} &=& \bar{T}_{31/32} \nonumber\\
        &=& \pm 2it^{13}_x \sin  k_{y/x} \pm 4it^{13}_{xy}\sin  k_{y/x} \cos  k_{x/y} \nonumber\\
        & & \mp 4it^{13}_{xxy}(\sin 2 k_{y/x} \cos  k_{x/y} - \cos 2 k_{x/y} \sin  k_{y/x}), \label{eq:T13} \\
T_{14/24} &=& \bar{T}_{41/42} \nonumber\\
        &=& 2it^{14}_x \sin  k_{x/y} + 4it^{14}_{xy}\sin  k_{x/y} \cos  k_{y/x} \nonumber\\
        & & + 4it^{14}_{xxy}\sin 2 k_{x/y} \cos  k_{y/x}, \label{eq:T14} \\
T_{15/25} &=& \bar{T}_{51/52} \nonumber\\
        &=& 2it^{15}_x \sin  k_{y/x} - 4it^{15}_{xy}\sin  k_{y/x} \cos  k_{x/y} \nonumber\\
        & & - 4it^{15}_{xxyy}\sin 2 k_{y/x} \cos 2 k_{x/y}, \label{eq:T15} \\
T_{34} &=& T_{43} \nonumber\\
      &=& 4t^{34}_{xxy}(\sin  k_x \sin 2 k_y - \sin 2 k_x \sin  k_y), \label{eq:T34} \\
T_{35} &=& T_{53} \nonumber\\
      &=& 2t^{35}_x(\cos  k_x - \cos  k_y) \nonumber\\
      & &+ 4t^{35}_{xxy}(\cos 2  k_x \cos  k_y - \cos  k_x \cos 2 k_y), \label{eq:T35} \\
T_{45} &=& T_{54} \nonumber\\
      &=& 4t^{45}_{xy}\sin  k_x\sin  k_y + 4t^{45}_{xxyy} \sin 2 k_x \sin 2 k_y. \label{eq:T45} 
\end{eqnarray}

\begin{table}
\caption{Parameters for the tight-binding portion of the five-orbital Model 1
  used here. The overall energy unit is
  electron volts.\label{tab:hopp5}} \centering
 \begin{tabular}{|c|ccccccc|}\hline
$t^{mn}_i$ & $i=x$ & $i=y$ & $i=xy$ & $i=xx$ & $i=xxy$ & $i=xyy$ & $i=xxyy$ \\
\hline
  $mn=11$  &$-0.355$ &$-0.17$ &$0.21$ &$-0.1$ & $0.01$ & $0$     & $0$       \\ 
\hline
  $mn=33$  &$0.1$   &       &$0.137$ &$-0.03$ &       &        &         \\ 
\hline
  $mn=44$  &$0.193$ &      &$-0.115$ &$0$    & $0$    &        & $0$       \\ 
\hline
  $mn=55$  &$-0.213$ &     &        &$0$    & $0$     &        & $0$       \\ 
\hline
  $mn=12$  &        &      &$-0.22$ &       & $0$     &        & $0$      \\ 
\hline
  $mn=13$  &$-0.35$ &      &$0.01$ &       & $0.02$    &        &        \\ 
\hline
  $mn=14$  &$0.55$ &      &$-0.13$ &       & $0.01$    &        &        \\ 
\hline
  $mn=15$  &$-0.25$ &      &$0$    &       &         &          & $0$    \\ 
\hline
  $mn=34$  &       &       &       &       & $-0.009$  &        &        \\ 
\hline
  $mn=35$  &$0.06$ &      &        &       & $-0.06$   &        &        \\ 
\hline
  $mn=45$  &       &      &$-0.05$ &       &          &        &  $0$     \\ 
\hline
 \end{tabular}

 \begin{tabular}{|cccc|}\hline
$\epsilon_{11}$ & $\epsilon_{33}$ & $\epsilon_{44}$ & $\epsilon_{55}$ \\
\hline
  $0.31$      &$-0.25$        &$0.43$          & $-0.8$       \\ 
\hline
 \end{tabular}
\end{table}




\begin{thebibliography}{99}

\bibitem{Fe-SC}
Y. Kamihara, T. Watanabe, M. Hirano, and H. Hosono, J. of the Am. Chem. Soc.
{\bf 130}, 3296 (2008).

\bibitem{chen1}
G.~F. Chen, Z. Li, G. Li, J. Zhou, D. Wu, J. Dong, W.~Z. Hu, P. Zheng, Z.~J.
Chen, H.~Q. Yuan, J. Singleton, J.~L. Luo, and N.~L. Wang, Phys. Rev. Lett.
{\bf 101}, 057007 (2008).

\bibitem{chen2}
G.~F. Chen, Z. Li, D. Wu, G. Li, W.~Z. Hu, J. Dong, P. Zheng, J.~L. Luo, and
N.~L. Wang, Phys. Rev. Lett. {\bf 100}, 247002 (2008).

\bibitem{chen3}
{X. H. Chen}, {T. Wu}, {G. Wu}, {R. H. Liu}, {H. Chen}, and {D. F. Fang},
Nature {\bf 453}, 761 (2008).

\bibitem{55}
Z.-A. Ren, W. Lu, J. Yang, W. Yi, X.-L. Shen, Z.-C. Li, G.-C. Che, X.-L. Dong,
L.-L. Sun, F. Zhou, and Z.-X. Zhao, Chin. Phys. Lett. {\bf 25}, 2215 (2008).


\bibitem{ren2}
Z.-A. Ren, G.-C. Che, X.-L. Dong, J. Yang, W. Lu, W. Yi, X.-L. Shen, Z.-C. Li,
L.-L. Sun, F. Zhou, and Z.-X. Zhao, EPL {\bf 83}, 17002 (2008).

\bibitem{david-j} D. C. Johnston, arXiv:1005.4392.

\bibitem{dai-lynn} J. W. Lynn and P. Dai, Physica C {\bf 469}, 469 (2009);
and references therein.

\bibitem{andy} M. D. Lumsden and A. D. Christianson, arXiv:1004.1969; and
references therein.

\bibitem{Cao08} C. Cao, P. J. Hirschfeld, and H.-P. Cheng, Phys.
Rev. B {\bf 77}, 220506(R) (2008); and references therein.

\bibitem{Raghu08} S. Raghu, X. L. Qi, C. X. Liu, D. J. Scalapino,
and S. C. Zhang, Phys. Rev. B {\bf 77}, 220503(R) (2008).

\bibitem{Kuroki} K. Kuroki, S. Onari, R. Arita, H. Usui, Y. Tanaka,
H. Kontani, and H. Aoki, Phys. Rev. Lett. {\bf 101}, 087004 (2008).

\bibitem{Daghofer08}
M. Daghofer, A. Moreo, J.~A. Riera, E. Arrigoni, D.~J. Scalapino, and E.
Dagotto, Phys. Rev. Lett. {\bf 101}, 237004 (2008).


\bibitem{Eremin} M. M. Korshunov and I. Eremin, Phys. Rev. B {\bf 78}, 140509(R)
(2008).

\bibitem{Si} Q. Si and E. Abrahams, Phys. Rev. Lett. {\bf 101}, 076401 (2008).

\bibitem{Hu} K. Seo, B. A. Bernevig, and J. Hu, Phys. Rev. Lett. {\bf 101}, 206404 (2008).

\bibitem{Lorenzana} J. Lorenzana, G. Seibold, C. Ortix, and M. Grilli, Phys. Rev. 
Lett. {\bf 101}, 186402 (2008).

\bibitem{Yuetal08} R. Yu, K. T. Trinh, A. Moreo, M. Daghofer,
J. A. Riera, S. Haas, E. Dagotto, Phys. Rev. B {\bf 79}, 104510 (2009).


\bibitem{Schmalian} R. Sknepnek, G. Samolyuk, Y.-bin Lee, and J. Schmalian, Phys. 
Rev. B {\bf 79}, 054511 (2009).


\bibitem{Moreo09} A. Moreo, M. Daghofer, J. A. Riera, and E. Dagotto,
Phys. Rev. B {\bf 79}, 134502 (2009).

\bibitem{Graser08} S. Graser, T. A. Maier, P. J. Hirschfeld, and D.
J. Scalapino, New J. Phys. {\bf 11}, 025016 (2009).

\bibitem{Graser10} A. F. Kemper, T. A. Maier, S. Graser, H-P. Cheng, P. J. Hirschfeld, and
D. J. Scalapino, New J. Phys. {\bf 12}, 073030 (2010).

\bibitem{chen102} W.-Q. Chen, K.-Y. Yang, Y. Zhou, and F.-C. Zhang, 
Phys. Rev. Lett. {\bf 102}, 047006 (2009).

\bibitem{calderon} 
M. J. Calder\'on, B. Valenzuela, 
and E. Bascones, New J. Phys. {\bf 11}, 013051 (2009).

\bibitem{calderon2} E. Bascones, 
M. J. Calder\'on, and B. Valenzuela, Phys. Rev. Lett. {\bf 104}, 227201 (2010).


\bibitem{wku} C.-C. Lee, W.-G. Yin, and W. Ku, Phys. Rev. Lett. {\bf 103}, 267001 (2009).

\bibitem{plee} P. A. Lee and X.-G. Wen, Phys. Rev. B {\bf 78}, 144517 (2008).

\bibitem{laad} M. S. Laad and L. Craco, Phys. Rev. Lett. {\bf 103}, 017002 (2009).

\bibitem{0910.2707} Sen Zhou and Ziqiang Wang, arXiv:0910.2707.

\bibitem{moreo09bis} 
A. Moreo, M. Daghofer, A. Nicholson, and E. Dagotto, Phys. Rev. B {\bf 80}, 104507 (2009).

\bibitem{ThreeOrbital}
M. Daghofer, A. Nicholson, A. Moreo, and E. Dagotto, Phys. Rev. B {\bf 81},
014511 (2010).

\bibitem{constraints} 
Xiaoyu Wang, Maria Daghofer, Andrew Nicholson, Adriana Moreo, Michael Guidry, and Elbio Dagotto,  
Phys. Rev. B {\bf 81}, 144509 (2010).

\bibitem{newpaper} 
M. Daghofer, Q.-L. Luo, R. Yu, D. X. Yao, A. Moreo, and E. Dagotto, Phys. Rev. B {\bf 81}, 
180514(R)(2010).

\bibitem{schrieffer} J. R. Schrieffer, X. G. Wen, and S. C. Zhang,
Phys. Rev. B {\bf 39}, 11663 (1989).

\bibitem{oles83}
A.~M. Ole{\'s}, Phys. Rev. B {\bf 28},  327  (1983).

\bibitem{photo2} L. X. Yang, Y. Zhang, H. W. Ou, J. F. Zhao, D. W. Shen, B. Zhou, J. Wei, 
F. Chen, M. Xu, C. He, Y. Chen, Z. D. Wang, X. F. Wang, T. Wu, G. Wu, X. H. Chen, M. Arita, K. Shimada, 
M. Taniguchi, Z. Y. Lu, T. Xiang, and D. L. Feng,  Phys. Rev. Lett. {\bf 102}, 107002 (2009).

\bibitem{photo3} Y. Zhang, J. Wei, H. W. Ou, J. F. Zhao, B. Zhou, F. Chen, M. Xu, C. He, 
G. Wu, H. Chen, M. Arita, K. Shimada, H. Namatame, M. Taniguchi, X. H. Chen, and D. L. Feng, 
Phys. Rev. Lett. {\bf 102}, 127003 (2009). 

\bibitem{photo5} 
M. Yi, D. H. Lu, J. G. Analytis, J.-H. Chu, S.-K. Mo, R.-H. He, M. Hashimoto, R. G. Moore, I. I. Mazin, D. J. Singh, Z. Hussain, I. R. Fisher, and Z.-X. Shen, Phys. Rev. B {\bf 80}, 174510 (2009).

\bibitem{photo6}
T. Shimojima, K. Ishizaka, Y. Ishida, N. Katayama, K. Ohgushi, T. Kiss, M. Okawa, T. Togashi, X. -Y. Wang, C. -T. Chen, S. Watanabe, R. Kadota, T. Oguchi, A. Chainani, and S. Shin, 
Phys. Rev. Lett. {\bf 104}, 057002 (2010).

\bibitem{photo1} T. Kondo, R. M. Fernandes, R. Khasanov, C. Liu, A. D. Palczewski, Ni Ni, M. Shi, A. Bostwick, E. Rotenberg, J. Schmalian, S. L. Bud'ko, 
P. C. Canfield, and A. Kaminski, Phys. Rev. B {\bf 81}, 060507(R) (2010).

\bibitem{photo7} S. de Jong, E. van Heumen, S. Thirupathaiah, R. Huisman, F. Massee, J. B. Goedkoop, R. Ovsyannikov, J. Fink, H. A. Duerr, A. Gloskovskii, H.S. Jeevan, P. Gegenwart, A. Erb, L. Patthey, M. Shi, R. Follath, A. Varykhalov, and M. S. Golde,  EPL {\bf 89}, 27007 (2010).

\bibitem{photo4} D. Hsieh, Y. Xia, L. Wray, D. Qian, K. Gomes, A. Yazdani, 
G.F. Chen, J.L. Luo, N.L. Wang, and M.Z. Hasan, arXiv:0812.2289.

\bibitem{y1} K. Hashimoto, T. Shibauchi, T. Kato, K. Ikada, R. Okazaki, H.
Shishido, M. Ishikado, H. Kito, A. Iyo, H. Eisaki, S. Shamoto,
and Y. Matsuda, Phys. Rev. Lett. {\bf 102}, 017002  (2009).

\bibitem{y2} T. Kondo, A. F. Santander-Syro, O. Copie, C. Liu, M. E.
Tillman, E. D. Mun, J. Schmalian, S. L. Budko, M. A. Tanatar,
P. C. Canfield, and A. Kaminski, Phys. Rev. Lett. {\bf 101}, 147003 (2008).

\bibitem{y3} H. Ding, P. Richard, K. Nakayama, K. Sugawara, T. Arakane, Y.
Sekiba, A. Takayama, S. Souma, T. Sato, T. Takahashi, Z. Wang,
X. Dai, Z. Fang, G. F. Chen, J. L. Luo, and N. L.Wang, EPL {\bf 83}, 47001 (2008).

\bibitem{y4} C. Martin, R. T. Gordon, M. A. Tanatar, M. D. Vannette, M. E.
Tillman, E. D. Mun, P. C. Canfield, V. G. Kogan, G. D.
Samolyuk, J. Schmalian, and R. Prozorov, arXiv:0807.0876.

\bibitem{y5} T. Y. Chen, Z. Tesanovic, R. H. Liu, X. H. Chen, and C. L.
Chien, Nature (London) {\bf 453}, 1224 (2008).

\bibitem{y6} D. Parker, O. V. Dolgov, M. M. Korshunov, A. A. Golubov, and
I. I. Mazin, Phys. Rev. B 78, 134524  (2008).





\bibitem{n1} L. Shan, Y. Wang, X. Zhu, G. Mu, L. Fang, C. Ren, and H.-H.
Wen, EPL {\bf 83}, 57004 (2008).

\bibitem{n2} M. Gang, Z. Xi-Yu, F. Lei, S. Lei, R. Cong, and W. Hai-Hu,
Chin. Phys. Lett. {\bf 25}, 2221 (2008).

\bibitem{n3} C. Ren, Z.-S. Wang, H. Yang, X. Zhu, L. Fang, G. Mu, L. Shan,
and H.-H. Wen, arXiv:0804.1726.

\bibitem{n4} K. Ahilan, F. L. Ning, T. Imai, A. S. Sefat, R. Jin, M. A.
McGuire, B. C. Sales, and D. Mandrus, Phys. Rev. B {\bf 78}, 100501(R) (2008).

\bibitem{n5} Y. Nakai, K. Ishida, Y. Kamihara, M. Hirano, and H. Hosono, J.
Phys. Soc. Jpn. {\bf 77}, 073701 (2008).

\bibitem{n6} H.-J. Grafe, D. Paar, G. Lang, N. J. Curro, G. Behr, J. Werner, J.
Hamann-Borrero, C. Hess, N. Leps, R. Klingeler, and B. B\"uchner,
Phys. Rev. Lett. {\bf 101}, 047003 (2008).

\bibitem{n7} K. Matano, Z. A. Ren, X. L. Dong, L. L. Sun, Z. X. Zhao, and
Guo-qing Zhen, EPL {\bf 83}, 57001 (2008).

\bibitem{n8} H. Mukuda, N. Terasaki, H. Kinouchi, M. Yashima, Y. Kitaoka,
S. Suzuki, S. Miyasaka, S. Tajima, K. Miyazawa, P. Shirage, H.
Kito, H. Eisaki, and A. Iyo, J. Phys. Soc. Jpn. {\bf 77}, 093704 (2008).

\bibitem{n9} O. Millo, I. Asulin, O. Yuli, I. Felner, Z.-A. Ren, X.-L. Shen,
G.-C. Che, and Z.-X. Zhao, Phys. Rev. B {\bf 78}, 092505 (2008).

\bibitem{n10} X. L. Wang, S. X. Dou, Z.-A. Ren, W. Yi, Z.-C. Li, Z.-X. Zhao,
and S.-I. Lee, arXiv:0808.3398. 

\bibitem{n11} K. Hashimoto, A. Serafin, S. Tonegawa, R. Katsumata, R. Okazaki, T. Saito, H. Fukazawa, Y. Kohori, K. Kihou, 
C. H. Lee, A. Iyo, H. Eisaki, H. Ikeda, Y. Matsuda, A. Carrington, and T. Shibauchi, arXiv:1003.6022, and references therein.

\bibitem{n12} B. Zeng, G. Mu, B. Shen, P. Cheng, H. Luo, H. Yang, L. Shan, C. Ren, and H-H. Wen,
arXiv:1006.2785.

\bibitem{manga} For a discussion of this relation in the manganite context
see E. Dagotto, T. Hotta, and A. Moreo, Phys. Rep. {\bf 344}, 1 (2001)
and references therein.

\bibitem{nomura} T. Nomura and K. Yamada, J. Phys. Soc. Jpn. {\bf 69}, 1856 (2000).

\bibitem{other-results} Other estimations of $U$ and $J$ have been presented in the literature.
In T. Miyake, K. Nakamura, R. Arita, and M. Imada, J. Phys. Soc. Jpn. {\bf 79}, 044705 (2010),
a combination of the constrained random-phase approximation and the maximally localized Wannier
function gives $J/U \sim 0.14$ and $U \sim 2.8$~eV. Other investigations,
such as those based on dynamical mean-field theory presented by H. Ishida 
and A. Liebsch, Phys. Rev. B {\bf 81}, 054513 (2010) also report a $U \sim 3$~eV, 
and $J/U \sim 0.25$. While these values of $J/U$ are at or close to the lower range of our 
``physical regions'', the value of $U$ differs from ours by approximately 
a factor 3 for the three-orbital
model and 2 for the five-orbital models. However, note that the
approaches cited above and ours are fundamentally different. In the first case, estimations for $J$
and $U$ are obtained directly based on ab-initio calculations, while in our approach it is
the comparison with experiments and the solution of the model via mean-field techniques that
are employed. The reason for this discrepancy in $U$ remains to be investigated. 

\bibitem{detail} Note that the results of Ref.~\onlinecite{Graser08} were obtained at a chemical potential
$\mu = 0$, which corresponds to a slightly different electronic density $n = 6.05$.

\bibitem{aoki:2009} K. Kuroki, H. Usui, S. Onari, R. Arita, and H. Aoki, Phys. Rev. B {\bf 79}, 224511 (2009).



\bibitem{berk-schrieffer} R. Berk and J.~R. Schrieffer, Phys. Rev. Lett. {\bf 14}, L369 (1966).

\bibitem{scalapino-hirsch:86} D.~J. Scalapino, E. Loh, and J. Hirsch, Phys. Rev. B {\bf 34}, 8190 (1986).

\bibitem{aguilar:1994} L. Puig-Puig, F. Lopez-Aguilar, and J. Costa-Quintana, Journ. Phys.-Cond. Matt. {\bf 6}, 4929 (1994).

\bibitem{quasi-nodes} B. Muschler, W. Prestel, R. Hackl, 
T. P. Devereaux, J. G. Analytis, Jiun-Haw Chu, and 
I. R. Fisher, Phys. Rev. B {\bf 80}, 180510 (2009). 

\bibitem{bulut} N. Bulut, D. J. Scalapino, and S. R. White, Phys. Rev. B {\bf 47}, 14599 (1993).

\bibitem{vojta} M. Vojta and E. Dagotto, Phys. Rev. B {\bf 59}, R713 (1999).



\end{thebibliography}
\end{document}